\documentclass[journal]{IEEEtran}
\usepackage{amsmath,amssymb,amsfonts}
\usepackage{amsthm}
\usepackage{calligra}
\usepackage[linesnumbered,ruled,vlined]{algorithm2e} 
\usepackage{bm}
\usepackage{cite}
\usepackage{hyperref}
\usepackage{enumerate}
\usepackage[table]{xcolor}
\usepackage{graphicx}
\usepackage{epstopdf}
\usepackage{cite,url,epsfig}
\usepackage{float}
\usepackage{caption}
\usepackage{comment}
\usepackage{booktabs}
\usepackage{subfigure}
\usepackage{url}
\usepackage[section]{placeins}
\usepackage{svg}
\usepackage{lipsum}


\begin{document}

\title{Multi-Agent DRL for Multi-Objective Twin Migration Routing with Workload Prediction in 6G-enabled IoV}

\author{
~Peng~Yin,~Wentao~Liang,~Jinbo~Wen,~Jiawen~Kang,~Junlong~Chen,~and Dusit~Niyato,~\IEEEmembership{Fellow, IEEE}
\thanks{P. Yin is with the Defence Industry Secrecy Examination and Certification Center (e-mail: pedro\_yin@163.com).

W. Liang is with the State Key Laboratory of Industrial Control Technology, College of Control Science and Engineering, Zhejiang University, Hangzhou, China (e-mail: WentaoL59@163.com).

J. Wen is with the College of Computer Science and Technology, Nanjing University of Aeronautics and Astronautics, China (e-mail: jinbo1608@nuaa.edu.cn).

J. Kang is with the School of Automation, Guangdong University of Technology, China (e-mail: kavinkang@gdut.edu.cn).

J. Chen is with the Intelligent Transportation Thrust, Hong Kong University of Science and Technology (Guangzhou), Guangzhou, China (e-mail: jchen241@connect.hkust-gz.edu.cn).

D. Niyato is with the College of Computing and Data Science, Nanyang Technological University, Singapore (e-mail: dniyato@ntu.edu.sg).

\textit{Corresponding author: Jiawen Kang}.}
}

\maketitle
\begin{abstract}
Sixth Generation (6G)-enabled Internet of Vehicles (IoV) facilitates efficient data synchronization through ultra-fast bandwidth and high-density connectivity, enabling the emergence of Vehicle Twins (VTs). As highly accurate replicas of vehicles, VTs can support intelligent vehicular applications for occupants in 6G-enabled IoV. Thanks to the full coverage capability of 6G, resource-constrained vehicles can offload VTs to edge servers, such as roadside units, unmanned aerial vehicles, and satellites, utilizing their computing and storage resources for VT construction and updates. However, communication between vehicles and edge servers with limited coverage is prone to interruptions due to the dynamic mobility of vehicles. Consequently, VTs must be migrated among edge servers to maintain uninterrupted and high-quality services for users. In this paper, we introduce a VT migration framework in 6G-enabled IoV. Specifically, we first propose a Long Short-Term Memory (LSTM)-based Transformer model to accurately predict long-term workloads of edge servers for migration decision-making. Then, we propose a Dynamic Mask Multi-Agent Proximal Policy Optimization (DM-MAPPO) algorithm to identify optimal migration routes in the highly complex environment of 6G-enabled IoV. Finally, we develop a practical platform to validate the effectiveness of the proposed scheme using real datasets. Simulation results demonstrate that the proposed DM-MAPPO algorithm significantly reduces migration latency by 20.82\% and packet loss by 75.07\% compared with traditional deep reinforcement learning algorithms.
\end{abstract}


\IEEEpeerreviewmaketitle

\section{Introduction}

\subsection{Background and Motivations}

{The Internet of Vehicles (IoV) enables network connections through vehicle-to-vehicle and vehicle-to-infrastructure communication, with mobile vehicles serving as information perception objects \cite{10034418,10852157}. To enhance traffic operation efficiency, vehicles need to acquire real-time information from their surrounding environments and infrastructure through built-in wireless sensors. The Sixth-Generation (6G) wireless communication technology is anticipated to achieve ultra-reliable and low-latency communications with large-scale coverage \cite{8869705}, supporting intelligent applications while meeting the reliability and security demands of future communication networks. With the adoption of 6G technology, the coverage and real-time performance of IoV will be significantly improved, encompassing connections across space, air, and ground \cite{9628162}. In 6G-enabled IoV networks, edge servers, such as RoadSide Units (RSUs), Unmanned Aerial Vehicles (UAVs), and satellites, can deploy computationally intensive vehicular applications, including 3D virtual human remote projection, augmented reality navigation, and autonomous driving \cite{9880528}, providing users with immersive and high-quality vehicular services \cite{9944868}.}

{Vehicle Twins (VTs), functioning as an integrated Digital Twin (DT) technology, are high-precision virtual replicas of vehicles that combine physical vehicle models, sensing data, and historical data \cite{kang2024hybrid}. VTs play a crucial role in 6G-enabled IoV by collecting and analyzing traffic data to monitor vehicle states and predict traffic conditions \cite{9745481}. However, constructing and updating VTs requires substantial computing and storage resources for real-time vehicle state monitoring and immersive service enhancement \cite{10185562}, and uploading directly to cloud servers will introduce additional communication latency. To address this, edge computing technology is introduced. Vehicles can offload VTs in the form of data packets to nearby edge servers (e.g., RSUs, UAVs, and satellites) after acquiring real-time data about themselves and their surroundings through multi-modal sensors. Then, vehicles with limited computing resources and energy reserves can be more effectively monitored for their status. Furthermore, with the support of 6G, vehicles can achieve reliable and ultra-low latency communications, ensuring real-time data updates and enhancing the responsiveness of vehicular applications.}

{Due to the dynamic mobility of vehicles and the uneven distribution of edge servers, the service range of a single edge server cannot fully cover the trajectory of vehicles in practical application scenarios, necessitating collaboration among multiple servers. Thus, it is challenging to ensure long-term stable communication interaction between a moving vehicle and its static VTs. To maintain the continuity of high-quality vehicular applications and services, VTs need to be dynamically migrated among edge servers based on the driving route of the corresponding vehicle. However, implementing effective VT migration presents several challenges:}

\begin{itemize}
    \item \textbf{Challenge \uppercase\expandafter{\romannumeral1}. Nonuniform deployment of ground edge servers:} The distribution of ground edge servers (i.e., RSUs) varies significantly across different regions. For example, in cities, RSUs are densely deployed and generally follow a more regular distribution pattern. In contrast, rural areas generally have a sparse and scattered distribution of edge servers, which may disrupt the continuity of VT migration.
    \item \textbf{Challenge \uppercase\expandafter{\romannumeral2}. Dynamic workload of edge servers:} The mobility of vehicles, along with their spatial and temporal uncertainties, causes the workload of edge servers to fluctuate dynamically. Overloaded edge servers can hinder VT migration efficiency, compromising the quality of services provided to users. Therefore, it is necessary to accurately predict the workload of edge servers during VT migration.
    \item \textbf{Challenge \uppercase\expandafter{\romannumeral3}. Calculation complexity of VT migration decision-making:} Deep Reinforcement Learning (DRL) algorithms are widely recognized for their effectiveness in identifying optimal decision-making strategies \cite{10638123}. However, traditional DRL algorithms may struggle to adapt to the complex and dynamic environment of VT migration, which may reduce the efficiency in identifying optimal VT migration routes.
\end{itemize}

\subsection{Solutions and Contributions}
{To address these challenges, we propose a novel workload prediction-based VT migration framework in 6G-enabled IoV. Specifically, we introduce UAVs, leveraging their high mobility to complement RSU computing resources on busy roads during peak hours, and incorporate Low-Earth-Orbit (LEO) satellites to provide vehicular services in remote areas. When terrestrial networks are overloaded, compromised, or unavailable, satellite communication can serve as a critical emergency solution, ensuring the continuity of core VT services \cite{9861699}. To prevent VT migration interruptions during peak traffic hours, we propose a Long Short-Term Memory (LSTM)-based Transformer model to accurately predict the long-term workload of edge servers, thus avoiding VT migration to overloaded edge servers. Furthermore, considering the complex environment of 6G-enabled IoV involving simultaneous multiple VT migrations, it is difficult for traditional DRL algorithms to efficiently identify optimal VT migration routes. To overcome this limitation, we propose a Dynamic Mask Multi-Agent Proximal Policy Optimization (DM-MAPPO) algorithm. The DM-MAPPO algorithm dynamically determines whether all output actions meet the requirements and reduces the probability of unreachable actions to zero. Finally, the action sequence obtained by DM-MAPPO represents the most efficient VT migration route\cite{9410247}. The main contributions of this paper are summarized as follows: }

\begin{itemize}
\item We propose a novel workload prediction-based VT migration framework in 6G-enabled IoV networks. This framework holistically incorporates RSUs, UAVs, and satellites as edge servers for VT migration, which can effectively address the problems of insufficient edge computing resources and low computing resource utilization in sparse regions (For Challenge \uppercase\expandafter{\romannumeral1}).
\item To accurately predict the long-term workload of edge servers, we propose an LSTM-based Transformer model that introduces the LSTM module into the Transformer architecture. By leveraging the proposed LSTM-based Transformer model, the workload sequence of edge servers can be effectively predicted and utilized for VT migration decision-making, thereby avoiding temporary interruption in VT migration during peak traffic hours (For Challenge \uppercase\expandafter{\romannumeral2}).
\item To efficiently obtain multiple optimal VT migration routes in the complex environment of 6G-enabled IoV networks, we propose a DM-MAPPO algorithm. Specifically, we integrate a dynamic mask module into the MAPPO algorithm, which sets the output probability of each invalid action to $0$, thereby dynamically reducing the action space and enhancing the efficiency of identifying optimal VT migration routes (For Challenge \uppercase\expandafter{\romannumeral3}).
\end{itemize}

The remainder of the paper is structured as follows: Section \ref{relatedwork} provides a review of the relevant literature. Section \ref{systemmodel} presents the workload prediction-based VT migration framework in 6G-enabled IoV networks. In Section \ref{problemformulation}, we formulate the optimization models for multiple VT migrations. In Section \ref{lstmbasedtransformer}, we propose the LSTM-based Transformer model to predict the workload of edge servers for VT migration. In Section \ref{madrl}, we propose the DM-MAPPO algorithm to identify multiple optimal VT migration routes. Section \ref{numericalresult} shows the simulation results of the proposed scheme. Finally, the paper is concluded in Section \ref{conclusion}.

\section{Related Work}\label{relatedwork}
\subsection{VTs in 6G-enabled IoV}
6G-enabled IoV is expected to become a key component of intelligent transportation systems \cite{10034418, 9806434, GUO2024237}. In \cite{10034418}, the authors highlighted the necessity of upgrading communication technology to support latency-sensitive applications in IoV. In 6G-enabled IoV, VTs are typically deployed on edge servers with sufficient computing resources to meet the low-latency requirements of vehicular services \cite{9737450, 10251544, ADNAN2024100615}. To alleviate the problem of excessive computational overhead caused by frequent VT updates, the authors in \cite{10551388} employed a double deep Q-network algorithm to make optimal offloading decisions. Similarly, in \cite{9726783}, the authors applied the deep deterministic policy gradient algorithm to deploy VTs on RSUs. In \cite{10185562}, the authors first predicted vehicle trajectories and utilized the Multi-Agent DRL (MADRL) algorithm to pre-migrate VTs to the target RSUs, thereby reducing the latency of subsequent VT migration. However, the above studies do not consider the uneven distribution of RSUs and the channel complexity of VT migration, which cannot guarantee the continuity of VT services. Therefore, it is necessary to introduce a new edge computing paradigm for VT migration in 6G-enabled IoV networks. 

\subsection{Workload Prediction of Edge Servers}
{To prevent VTs from being migrated to overloaded edge servers, it is essential to predict the long-term workload of edge servers in advance. In the field of time series prediction, particularly in traffic flow forecasting, traditional algorithms often struggle to deliver superior performance due to the complex temporal dependencies, high volatility, and nonlinear characteristics of traffic flow data. Fortunately, the emergence of Recurrent Neural Networks (RNNs) has made precise prediction of long-term time series data possible \cite{chen2023long}. Gated Recurrent Unit (GRU) and LSTM networks retain long-term memory through gating mechanisms and cell states, offering both effectiveness and lightweight properties \cite{10288593, SUN2023105662}. Based on the combination of graph neural networks and RNNs, the authors in \cite{10.1145/3532611} proposed a dynamic graph convolutional cycle network to improve traffic prediction accuracy by capturing dynamic correlation features between vehicle locations in the road network. In recent years, more prediction algorithms have been improved on the basis of Transformers and carried out long-term time series prediction\cite{ZHOU2023103886}. For instance, the authors in \cite{liu2022pyraformer} proposed the Pyraformer algorithm, which captures multi-resolution features through an inter-scale tree structure. In \cite{Zhou_Zhang_Peng_Zhang_Li_Xiong_Zhang_2021}, the authors proposed the Informer algorithm, which utilizes a prob-sparse self-attention mechanism with a distillation technique to efficiently extract key information. Since the strong periodicity of traffic flow data, these algorithms generally focus on capturing long-term periodicity information while neglecting some short-term time series variations. However, edge servers are highly sensitive to fluctuations in traffic flow data, making accurate workload prediction crucial for efficient VT migration decisions.}

\begin{figure}[t]
    \centering
    \includegraphics[width=0.48\textwidth]{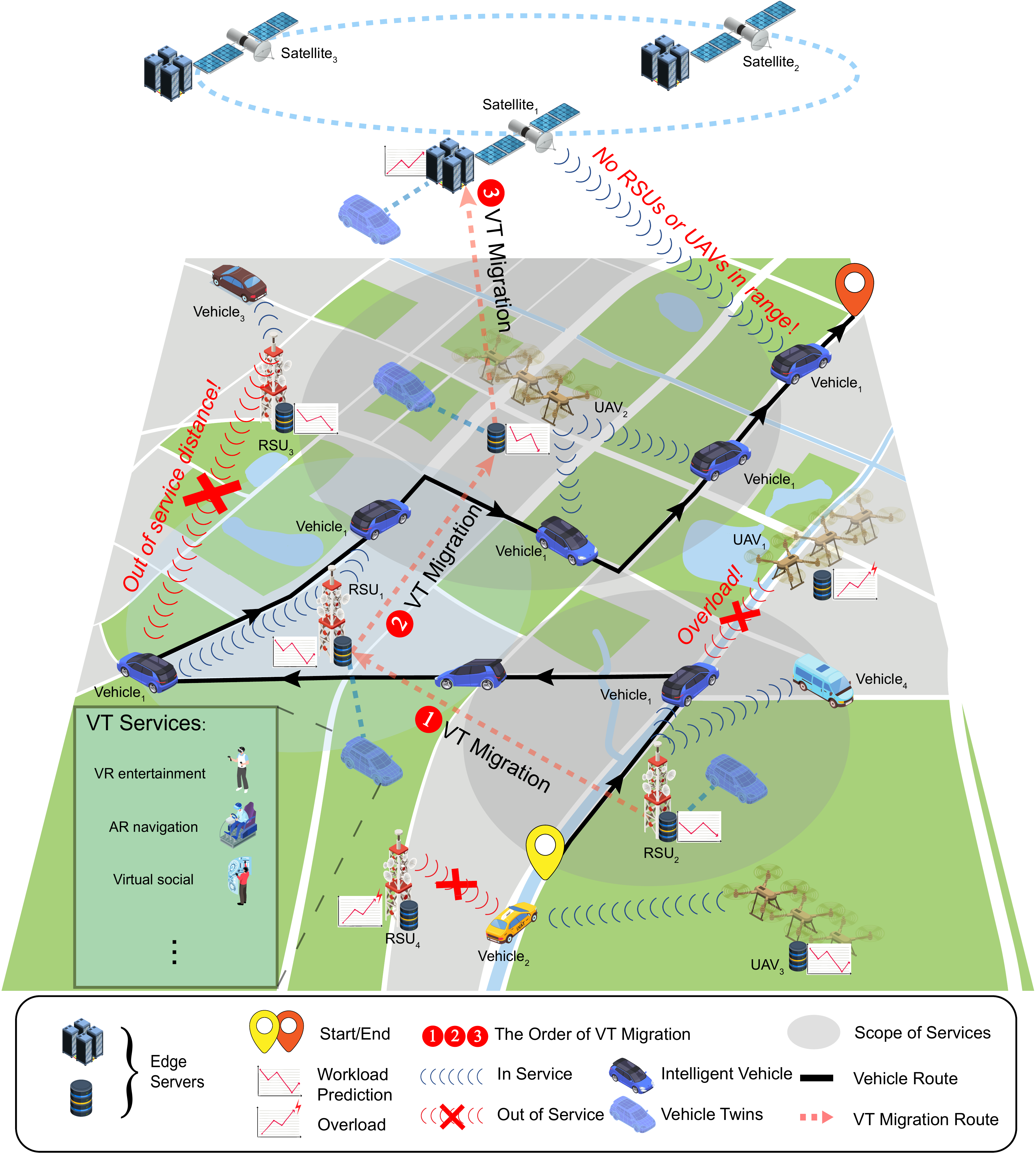}
    \caption{The workload prediction-based VT migration framework in 6G-enabled IoV networks.}
    \label{General_view}
\end{figure}

\subsection{MADRL for Routing Tasks}
MADRL algorithms are powerful tools for obtaining optimal decision strategies in routing tasks, enabling collaborative task planning and multi-objective strategy optimization among DRL agents. In \cite{10526476}, the authors utilized collaborative MADRL algorithms to optimize routing policies, which demonstrates the advantages of MADRL in task planning and strategy optimization. Among various MADRL algorithms, MAPPO stands out as an efficient and stable optimization algorithm, which is widely utilized in routing tasks \cite{10026882}. For example, the authors in \cite{10026882} utilized the MAPPO algorithm to assign mobile policies to each UAV, thereby maximizing the service quality for ground equipment. In \cite{10572232}, the authors adopted the MAPPO algorithm to select edge service migration targets in the multi-reconfigurable intelligent surface-assisted vehicular edge computing network. However, the actions selected by MAPPO may not always comply with physical constraints. Specifically, some actions attempt to migrate VTs to edge servers that are either out of service or overloaded, potentially leading to vehicular service outages. In addition, the number of edge servers is positively correlated with the action space of MAPPO. As the number of edge servers increases, the likelihood of selecting invalid actions also grows. Moreover, the training time of MAPPO is closely related to the size of the action space. Notably, the authors in \cite{zhong2024no} proved that in environments with large action spaces, the operation of trying all possible actions at each step is redundant, deteriorating algorithm convergence and reducing the long-term rewards obtained. Since VT migration routing contains a large action space, it is necessary to design an efficient MADRL algorithm capable of identifying optimal VT migration routes.

\section{System Model}\label{systemmodel}
As shown in Fig. \ref{General_view}, we propose a workload prediction-based VT migration framework in 6G-enabled IoV networks, consisting of $V$ vehicles, $N$ RSUs, $U$ UAVs, and one satellite. The framework is simulated over multiple time slots. In each time slot $t$, we denote the vehicle set and the RSU set as $\mathcal{V}^t=\{1,\ldots,v ,\ldots,V \}$ and $\mathcal{N}_v^t=\{1,\ldots, n_v^t,\ldots, N \}$, respectively, where $n_v^t$ represents the RSU $n$ providing vehicular services to vehicle $v$ in time slot $t$. The UAV set is denoted as $\mathcal{U}_v^t=\{1, \ldots, u_v^t, \ldots, U\}$, where $u_v^t$ represents the UAV $u$ serving vehicle $v$ in time slot $t$. Additionally, we denote the satellite providing vehicle $v$ in time slot $t$ as $s_v^t$. Thus, the total number of edge servers in this framework is $M = N+U+1$, where the edge server set is given by $\mathcal{M}_v^t=\mathcal{N}_v^t \cup \mathcal{U}_v^t \cup s_v^t = \{1,\ldots,m_v^t,\ldots, M \}$, including the RSU set, the UAV set, and the satellite. The maximum communication distance of edge server $m$ is $D_m^{max}$, and the maximum cache size is $L_m^{max}$.

\subsection{Network Model}
{
In time slot $t$, the latitude and longitude of vehicle $v$ are denoted as $\phi_v^t$ and $\lambda_v^t$, respectively, while the latitude and longitude of edge server $m$ are represented as $\phi_m^t$ and $\lambda_m^t$, respectively. We denote the latitude difference between vehicle $v$ and edge server $m$ as $\Delta\phi_{v,m}^t=\frac{\pi}{180}(\phi_v^t-\phi_m^t)$, and the longitude difference as $\Delta\lambda_{v,m}^t=\frac{\pi}{180}(\lambda_v^t-\lambda_m^t)$. The conversion factor from degrees to radians $\vartheta$ is defined as $\vartheta=\frac{\pi}{180}$. Based on the Haversine function, the central angular radian $\eta_{v,m}^t$ corresponding to the distance $D_{v,m}(t)$ between vehicle $v$ and edge server $m$ can be calculated as}
{
\begin{equation}
    \eta_{v,m}^t =2 \arcsin{\sqrt{\mathrm{hav}(\eta_{v,m}^t)}},
\end{equation}
where the Haversine function is $\mathrm{hav}(\eta_{v,m}^t) = \sin^2\Big(\frac{\eta_{v,m}^t}{2}\Big) =\sin^2 (\Delta\lambda_{v,m}^t) +\cos(\vartheta\phi_{v}^{t})\cos(\vartheta\phi_{m}^{t}) \sin^2 (\Delta\phi_{v,m}^t)$.}

Considering the Earth's curvature and the altitude difference between UAVs and the ground, the distance $D_{v,m}(t)$ between vehicle $v$ and edge server $m$ in time slot $t$ is given by
\begin{equation}
    \label{D_v_m}
    D_{v,m}(t)=
    \begin{cases} 
        R \cdot \eta_{v,m}^t, &  m \in \mathcal{N}_v^t, \\
        \sqrt{(R \cdot \eta_{v,m}^t)^2+h_u^2}, &  m \in \mathcal{U}_v^t,
    \end{cases}
\end{equation}
where $h_u$ represents the cruise altitude of UAV $u$, and $R$ represents the radius of the Earth, standing at $6,371.393\:\rm{km}$. Similarly, the distance $D_{m,m_d}^{mig}(t)$ between edge server $m$ and edge server $m_d \in \mathcal{M}_v^t$ can be calculated using Eq. (\ref{D_v_m}) based on $\eta_{m,m_d}^t,\: m_d \in \mathcal{M}_v^t$.

Considering homogeneous channels between vehicles and edge servers, the gain of the Rayleigh fading channel between vehicle $v$ and edge server $m$ is given by \cite{10185562}
\begin{equation}
    h_{v,m}(t)=A \left( \frac{c}{4\pi f D_{v,m}(t)} \right)^2,
    \label{Rayleigh}
\end{equation}
where $c=3 \times 10^8\:\rm{m/s}$ represents the speed of light, $A$ represents the channel increment coefficient, and $f$ represents the carrier frequency.

We consider that the orthogonal frequency division multiplexing access technology is applied in the proposed framework \cite{10271832}, indicating that the communication channels between different vehicles and an edge server are orthogonal. Based on Eq. (\ref{Rayleigh}), the uplink transmission speed between vehicle $v$ and edge server $m$ can be calculated as\cite{6773024}
\begin{equation}\label{uplink_transmission}
    R_{v,m}^{up}(t)=B_m^{up}\log{\left( 1+\frac{p_vh_{v,m}(t)}{\sigma^2} \right)},
\end{equation}
where $\sigma$ represents the additive white Gaussian noise, $p_v$ represents the transmit power of vehicle $v$, and $B_m^{up}$ represents the uplink bandwidth.

Similarly, the downlink transmission speed between vehicle $v$ and edge server $m$ can be calculated as
\begin{equation}
    R_{v,m}^{down}(t)=B_m^{down}\log{\left( 1+\frac{p_m h_{v,m}(t)}{\sigma^2} \right)},
\end{equation}
where $p_m$ represents the signal transmission power of edge server $m$, and $B_m^{down}$ represents the downlink bandwidth.

Satellites are generally considered to be in geosynchronous orbit at $35,786\:\rm{km}$ above the Earth's surface, while LEO satellites typically operate at an altitude of $550\:\rm{km}$. In each time slot, the size of the data generated by vehicle $v$, including VT migration requests, is denoted as $S_v$. Based on the uplink transmission speed, the uplink latency $T_{v,m}^{up}(t)$ is given by
\begin{equation}
T_{v,m}^{up}(t)=
\begin{cases} 
\frac{S_v}{R_{v,m}^{up}(t)}, &  m \neq s_v^t, \\
\frac{S_v}{R^{SAT}}, &  m = s_v^t,
\end{cases}
\end{equation}
where $R^{SAT}$ is the Starlink data transmission speed\footnote{Data from Starlink website: \url{https://www.starlink.com/technology}}.
Similarly, the downlink latency $T_{v,m}^{down}(t)$ can be calculated as
\begin{equation}
T_{v,m}^{down}(t)=
\begin{cases} 
\frac{S_v}{R_{v,m}^{down}(t)}, &  m \neq s_v^t, \\
\frac{S_v}{R^{SAT}}, &  m = s_v^t.
\end{cases}
\end{equation}

\subsection{Computing Model}
{Similar to RSUs, UAVs are considered to have certain computational capabilities \cite{7932157}, and we define the size of uncompleted tasks in the cache of the edge server $m$ in time slot $t$ as $S_m(t)$. Thus, the load of edge server $m$ in time slot $t$ can be calculated as}
\begin{subequations}
    \begin{align}
        S_m'(t)&=S_m(t)+\sum_{v=1}^{V}{g_{v,k}S_v},\\
        g_{v,k}&=
        \begin{cases} 
            0, &  k \neq m^t, \\
            1, &  k = m^t,
        \end{cases}
    \end{align}
\end{subequations}
where $g_{v,k} = 1$ indicates that vehicle $v$ is served by edge server $m$, otherwise $g_{v,k} = 0$. When the amount of data to be processed by the edge server $m$ in time slot $t$ becomes too large, it may cause data packet dropout during VT migration, affecting the vehicular services provided by VTs to users. If RSUs or UAVs become overloaded, the VTs need to be offloaded to the satellite to obtain additional computing resources, albeit at a higher cost. Thus, the computing latency is given by \cite{9300168}
\begin{equation}
    T_{v,m}^{cul}(t)=
    \begin{cases} 
        \frac{e_v S_m'(t)}{C_m}, &  S_m'(t) \le L_m^{max}, \\
        \frac{e_v S_v}{C_s}, &  S_m'(t) > L_m^{max},
    \end{cases}
\end{equation}
where $e_v$ represents the number of computation cycles required per bit, $C_m$ denotes the Central Processing Unit (CPU) operating frequency of edge server $m,\: m \neq s_v^t$, and $C_s$ denotes the CPU operating frequency of the satellite.

\subsection{Migration Model}
In this framework, VTs need to be migrated among edge servers to ensure short-distance communication with vehicles. However, VT migration will increase the workload of edge servers, leading to problems such as time alignment, resource waste, and service interruption. If VT migration occurs at edge server $m$, the state of edge server $m$ in time slot $t$ will be different from the state of time slot $t+1$. Thus, whether VT migration can occur at edge server $m$ can be judged by
\begin{equation}
    \label{jump_Eq}
    J_m^t=
    \begin{cases} 
        1, &  m^t \neq m^{t+1}, \\
        0, &  m^t = m^{t+1}.
    \end{cases}
\end{equation}

After edge server $m$ receives a VT migration request from vehicle $v$, the VTs of vehicle $v$ are migrated from the originally connected edge server $m$ to the target edge server $m_d$, and the migration transmission speed can be calculated as
\begin{equation}
    \label{Rmig}
    R_{m,m_d}^{mig}(t)=B_{m,m_d}^{mig}\log{\left( 1+\frac{p_m h_{m,m_d}^{mig}(t)}{\sigma^2} \right)},
\end{equation}
where $B_{m,m_d}^{mig}$ represents the migration transmission bandwidth, $p_m$ represents the transmit power of edge server $m$, and $h_{m,m_d}^{mig}(t) = A \big( \frac{c}{4\pi f D^{mig}_{m,m_d}(t)} \big)^2$.

Based on Eq. (\ref{jump_Eq}) and Eq. (\ref{Rmig}), the cost of VT migration can be quantified as the transmission latency $T_{v,m,m_d}^{mig}(t)$ between edge server $m$ and edge server $m_d$, given by\cite{10185562}
\begin{equation}
    T_{v,m,m_d}^{mig}(t)=J_m^t \frac{S_v^{mig}}{R_{m,m_d}^{mig}(t)},
\end{equation}
where $S_v^{mig}$ represents the size of migrated VT data corresponding to the vehicle $v$.

Thus, the total latency $T_{v,m}^{sum}(t)$ of VT migration from edge server $m$ to edge server $m_d$ is given by
\begin{equation}
    T_{v,m}^{sum}(t)=T_{v,m}^{up}(t)+T_{v,m}^{down}(t)+T_{v,m}^{cul}(t)+T_{v,m,m_d}^{mig}(t).
\end{equation}

{If vehicle $v$ is out of the service range of the connected edge server $m$, or the connected edge server $m$ is overloaded, the VT service will be interrupted. Thus, whether the VT service of vehicle $v$ fails in time slot $t$ is given by}
\begin{equation}
    \label{drop_eq}
    \xi_v^t=
    \begin{cases} 
        1, &  D_{v,m}(t) > D_m^{max} \ \text{or} \ S_m'(t)>L_m^{max},  \\
        0, &  \text{otherwise},
    \end{cases}
\end{equation}
where $\xi_v^t = 1$ means that the VT migration of vehicle $v$ fails.

In addition, vehicle $v$ begins driving in time slot $t = 0$ and reaches its destination in time slot $t_v^{end}$. Based on Eq. (\ref{drop_eq}), the packet loss rate during VT migration of vehicle $v$ is denoted as the ratio of the number of packet losses to the total number of time slots, which can be calculated as
\begin{equation}
    q_v=\frac{\sum_{t=0}^{t_v^{end}}{\xi_v^t}}{t_v^{end}} \times \text{100\%}.
\end{equation}

\section{Problem Formulation}\label{problemformulation}
{In this paper, we aim to optimize three objectives. The first goal is to minimize the total latency $T_{v,m}^{sum}(t)$ of all vehicles, thus ensuring a fully immersive experience. Secondly, the uneven distribution of RSUs in a region may cause cyclical and regional load peaks of edge servers. To balance the workload of all edge servers in the region, we take the variance of the workload of all edge servers $\mathrm{Var}(S_m'(t))$ as one of the optimization objectives, and $\mathrm{Var}(S_m'(t))$ can be calculated as $\mathrm{Var}(S_m'(t))=\frac{1}{M}\sum_{i=1}^{M}{(S_i'(t)-\mu _i(t))^2}$, where $\mu_i(t)$ represents the average of $S_m'(t)$. Finally, frequent VT migrations not only lead to additional unnecessary computing overhead but also degrade the quality of vehicular services. Thus, the goal is to minimize the total number of VT migrations after optimization, and the three optimization objectives along with their corresponding constraints are formulated as}
\begin{subequations}\label{formulated_problem}
    \begin{align}
        \min \quad & O_T=\sum_{v=1}^{V}{\sum_{t=0}^{t_v^{end}}{T_{v,m}^{sum}(t)}},\label{O_T} \\
        \min \quad & O_V=\sum_{t=0}^{t_v^{end}}{\mathrm{Var}(S_m'(t)) },\label{O_V} \\
        \min \quad & O_D=\sum_{v=1}^{V}{q_v}, \label{O_D}\\
        \rm{s.t.} \quad 
            & S_m'(t) \le L_m^{max},\label{con1} \\
            & D_{v,m}(t) \le D_m^{max},\label{con2} \\
            & m \in \mathcal{M}_v^t ,\label{con3}
    \end{align}
\end{subequations}
where $O_T$ represents the total latency, $O_V$ represents the total workload variance, and $O_D$ represents the total packet loss rate. Constraint (\ref{con1}) is set to prevent packet loss or disconnection when VT data sent to the edge server $m$ exceeds the maximum cache. Constraint (\ref{con2}) is set to prevent connected edge servers from going out of communication range. Constraint (\ref{con3}) is set to prevent connections to edge servers that are not in the region.

Accurate workload data is crucial for VT migration planning tasks. Due to the large number of VT migration targets and the high communication and computing resource requirements of VT tasks, a large prediction error will lead to the failure of VT migration performed by the DRL algorithm, reducing the quality of the overall VT migration route. 

\section{LSTM-based Transformer Model for Workload Predictions of Edge Servers}\label{lstmbasedtransformer}

\begin{figure}[t]
    \centering
    \includegraphics[width=0.45\textwidth]{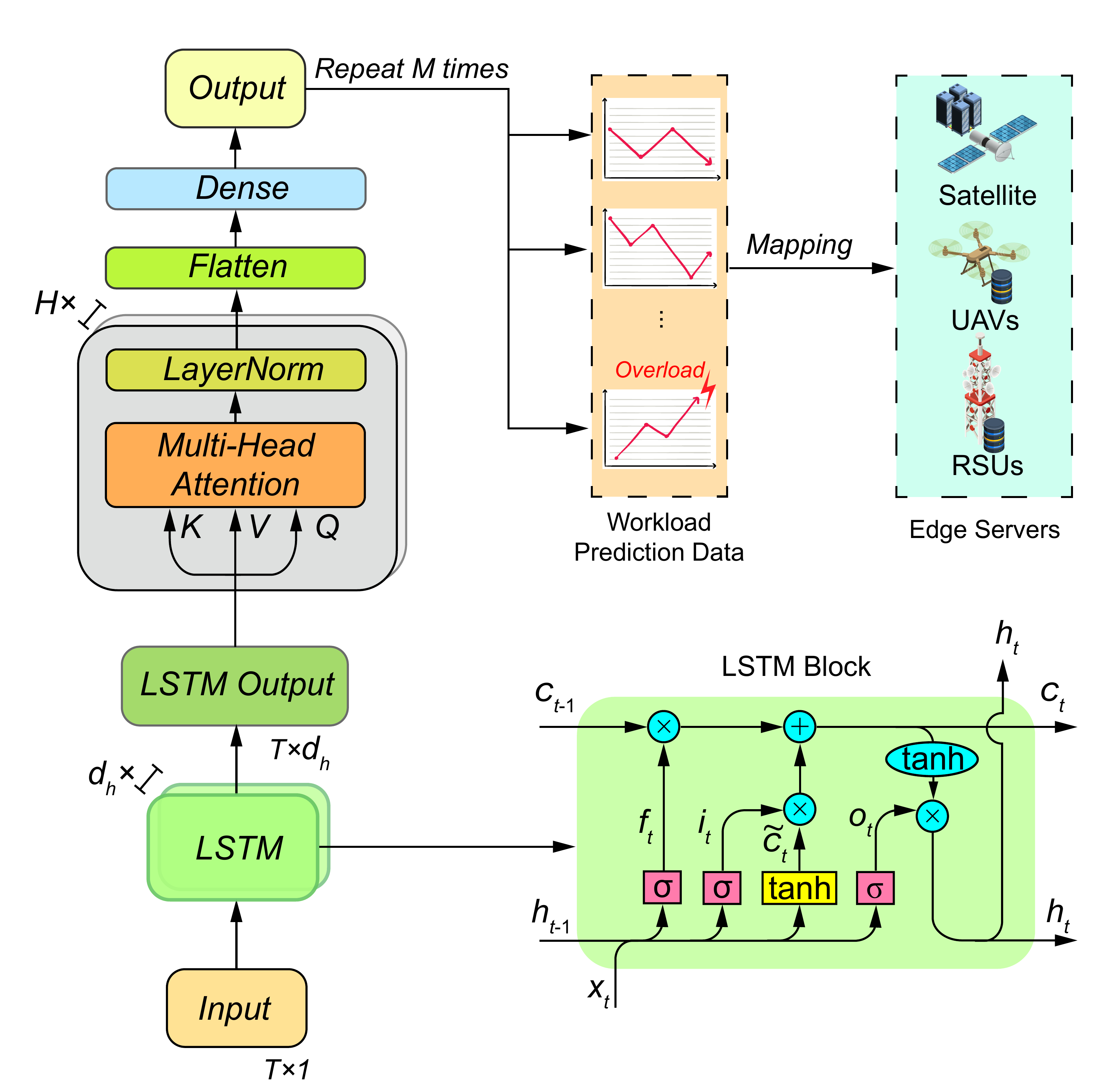}
    \caption{The structure of the LSTM-based Transformer model.}
    \label{LSTMTranstructure}
\end{figure}

{The workload of edge servers, while exhibiting periodic patterns \cite{YADAV2024122333}, presents significant prediction challenges due to its complex temporal dependencies, high volatility, and nonlinear characteristics. LSTM typically performs well in predicting periodic data \cite{10233845}, but it performs poorly for aperiodic data, especially in capturing fine-grained workload variations. In contrast, the Transformer model can effectively capture the relationships among vectors, making it a key technique for generating accurate contextual semantics in natural language processing \cite{vaswani2017attention}. In this section, we propose the LSTM-based Transformer model to predict the workload of edge servers. The structure of the proposed model is shown in Fig. \ref{LSTMTranstructure}. }

\begin{algorithm}[t]
\DontPrintSemicolon
\SetAlgoLined
    \caption{LSTM-based Transformer Model}\label{LSTM-Tran}
    \KwIn{The sequence of traffic flow data $\mathrm{X}_m$.}
    \KwOut{$\rm{Model}_m$, the MSE loss $L_m$ of $\rm{Model}_m$, predicted workload sequences of each edge server $m$.}
    \textbf{\textit{Phase 1: Initialization} }\\
        Initialize the traffic flow prediction model $\rm{Model}_m$. The data $\mathrm{X}_m$ are preprocessed as $S_m(t),\ldots,S_m(t+T)\rightarrow S_m(t+T+1)$.\\
    \textbf{\textit{Phase 2: Training}} \\
        \For{$t=0,1,\ldots,t_{end}-T$}
            {
                Input historical traffic flow data $\mathrm{X}_m'=[S_m(t),\ldots,S_m(t+T)]$.\\
                Obtain predicted traffic flow data $S_m^{pre}(t+T+1)$.\\
                Calculate the MSE loss $L_m=(S_m^{pre}(t+T+1)-S_m(t+T+1))^2$.
            }
    \textbf{\textit{Phase 3: Evaluation}} \\
        Input historical traffic flow data $\mathrm{X}_m'$ into the trained traffic flow prediction model $\rm{Model}_m$.\\
        Obtain predicted traffic flow data $S_m^{pre}(t+T+1)$ through $\rm{Model}_m$. \\
        Calculate the accuracy rate ${Acc}_m=S_m^{pre}(t+T+1)-S_m(t+T+1)$ for each edge server.\\
        \For{$m=1,2,\ldots,M-1$}
            {
                Deploying the trained model $\rm{Model}_m$ to the edge server $m$. \\
                Input historical traffic flow data $\mathrm{X}_m'$ into edge server $m$ to obtain the predicted traffic flow data $S_m^{pre}(t)$.\\
                Calculate the predicted workload of edge server $m$ as $S_m'(t)=S_m(t)+S_v, \: m=m_v^t$.\\
            }
\end{algorithm}

The input time-series workload data\footnote{RSU coordinate dataset: \url{https://pems.dot.ca.gov/}} with length $T$ is first processed through each LSTM module to capture the long-term cycle dependency of each edge server. Then, the output of each LSTM module $\mathbf{H}_t=[h_t, h_{t+1}, \ldots, h_{t+T}]$ will be transformed into data with shape $T \times d_k$ through the dense layer, where $d_k$ represents the vector dimension used by the query and key vectors in the self-attention mechanism. Finally, the processed data is utilized as input to the Transformer module to predict the workload data of edge servers accurately. The prediction result will be mapped to the corresponding edge server, and the amount of data accumulated in the cache of the edge server will be calculated. The structure of the LSTM module is given by \cite{GULMEZ2023120346}
\begin{subequations}
    \begin{align}
        \ i_t&=\sigma{(x_t \mathbf{W}^i_x + h_{t-1} \mathbf{W}^i_h+\mathbf{b}_i)},\\
        \ f_t&=\sigma{(x_t \mathbf{W}^f_x + h_{t-1} \mathbf{W}^f_h+\mathbf{b}_f)},\\
        \ o_t&=\sigma{(x_t \mathbf{W}^o_x + h_{t-1} \mathbf{W}^o_h+\mathbf{b}_o)},\\
        \ \widetilde{C_t}&=\tanh(x_t \mathbf{W}^c_x + h_{t-1} \mathbf{W}^c_h+\mathbf{b}_c),\\
        \ C_t&=\sigma{(f_t\cdot C_{t-1}+i_t\cdot \widetilde{C_t})},\\
        \ h_t&=o_t\cdot\tanh(C_t),
    \end{align}
\end{subequations}
where $i_t$ represents the output of the input gate, $f_t$ represents the output of the forget gate, $o_t$ represents the output of the output gate, $h_t$ represents the hidden state, $C_t$ represents the cell state, $\widetilde{C_t}$ represents the candidate cell state, $\mathbf{W}^i_h$, $\mathbf{W}^f_h$, $\mathbf{W}^o_h$, $\mathbf{W}^c_h$, $\mathbf{W}^i_x$, $\mathbf{W}^f_x$, $\mathbf{W}^o_x$, and $\mathbf{W}^c_x$ represent the weight matrices of gates, and $\mathbf{b}_i$, $\mathbf{b}_f$, $\mathbf{b}_o$, and $\mathbf{b}_c$ represent the bias term of gates.

The Transformer module employs the multi-head attention mechanism to process the input data. $\mathbf{Q}_t^h$, $\mathbf{K}_t^h$, and $\mathbf{V}_t^h$ represent the query, key, and value vectors, respectively. Thus, the structure of the Transformer module is given by \cite{vaswani2017attention} 
\begin{subequations}
    \begin{align}
        \mathbf{Q}_t^h&=\mathbf{H}_t \mathbf{W}_Q,\\
        \mathbf{K}_t^h&=\mathbf{H}_t \mathbf{W}_K,\\
        \mathbf{V}_t^h&=\mathbf{H}_t \mathbf{W}_V,\\
        \begin{split}
        \rm{head}_h &=
         \mathrm{softmax}\left(  \frac{\mathbf{Q}_t^h {\mathbf{K}_t^h}^T}{\sqrt{d_k}} \right),
         \end{split}\\
        \rm{Output}&=\mathrm{Concat}(\mathrm{head}_1,\ldots,\mathrm{head}_1)\mathbf{W}_O,
    \end{align}
\end{subequations}
where $\mathbf{W}_Q$, $\mathbf{W}_K$, $\mathbf{W}_V$, and $\mathbf{W}_O$ represent the trainable parameter matrices.

The output of the Transformer module represents the traffic flow $c_m(t)$ of each edge server, which can be mapped to the workload data of each edge server through $S_m(t)=S_v \cdot c_m(t)$. Algorithm \ref{LSTM-Tran} presents the pseudocode of the LSTM-based Transformer model, whose computational complexity is $\mathcal{O}(T \times M \times (H \times T^2 \times d_k+T \times d_h))$, where $H$ denotes the number of attention heads, and $d_h$ denotes the hidden dimension of each LSTM module. By leveraging the proposed LSTM-based Transformer model, VTs can avoid being migrated to overloaded edge servers. Similar to the traveling salesman problem \cite{cook2023complexity}, the formulated problem (\ref{formulated_problem}) is NP-hard, with a time complexity of $\mathcal{O}(n^n)$, making it nearly impossible to find the optimal solution through enumeration within a finite time. Fortunately, DRL algorithms provide solutions to NP-hard problems. However, traditional DRL algorithms suffer from long training times, leading to low efficiency and suboptimal solutions. Thus, we propose the DM-MAPPO algorithm to effectively solve the problem (\ref{formulated_problem}).

\section{DM-MAPPO Algorithms for Optimal VT Migration Routes}\label{madrl}

In this section, we first model the VT migration routing task as a Partially Observable Markov Decision Process (POMDP). Then, we introduce the DM-MAPPO algorithm.

\begin{figure}[t]
\centering
\includegraphics[width=0.48\textwidth]{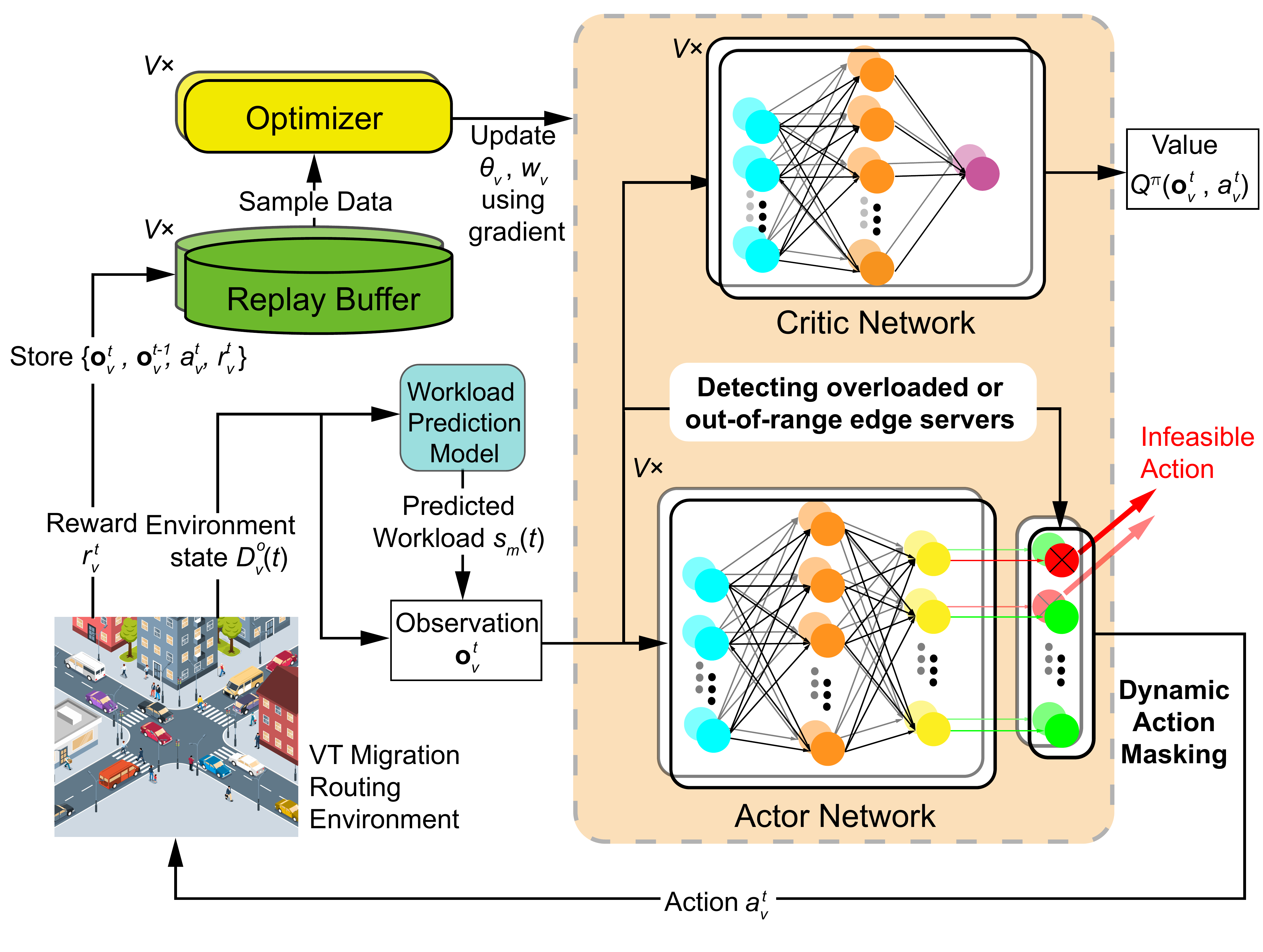}
\caption{The algorithm architecture of DM-MAPPO for optimal VT migration routes.
}
\label{MAPPO}
\end{figure}

\subsection{POMDP for VT Migration Routing Tasks}
\begin{itemize}
    \item \textbf{Observation Space:} 
    During VT migration, vehicles can observe real-time information from the surrounding environment. The distance set of vehicle $v$ from edge servers is denoted as $\mathbf{D}_v^o(t)=\{ D_{v,1}(t),\ldots,D_{v,m}(t),\ldots,D_{v,M-1}(t) \}$. In this framework, the number of satellites is $1$, and the communication distance and transmission rate have been determined. Thus, the distance between the satellite and the vehicle is not added to the observation set. The workload prediction set is denoted as $\mathbf{S}^o(t)=\{S_1^{pre}(t),\ldots,S_m^{pre}(t)\ldots,S_M^{pre}(t)\}$. For vehicle $v$, its observation in time slot $t$ is denoted as
    \begin{equation}
        \mathbf{o}_v^t=[\mathbf{D}_v^o(t),\mathbf{S}^o(t)].
    \end{equation}

    Remarkably, all vehicles in the same area observe $\mathbf{S}^o(t)$ in a shared way. 

    \item \textbf{Action Space:} The action of each DRL agent is to select the edge server that provides vehicular services for vehicle $v$ in time slot $t$. Thus, the action space can be denoted as $\mathcal{A}_v^t=[a_{v,1}^t,\ldots, a_{v,m}^t,\ldots,a_{v,M}^t]$, where $a_{v,m}^t$ presents that the DRL agent migrates VTs of vehicle $v$ to edge server $m$. Beside, we denote the actor network output for each DRL agent as $\mathbf{P}_v^t=[P_{v,1}^t,\ldots, P_{v,m}^t,\ldots, P_{v,m}^t]$, where $\sum_{i=1}^{M}{P_{v, i}^t=1}$. Finally, we perform the action $a_v^t$ with the maximum probability as
    \begin{equation}
        a_v^t=\underset{m}{\mathrm{argmax}} (\mathbf{P}_v^t).
    \end{equation}
    
    Since VTs can only exist on one edge server in each time slot $t$,  the sampled action $a_v^t$ is one-dimensional.
    
    \item \textbf{Immediate Reward:} The weighted summation method is utilized to transform the multi-objective optimization problem (\ref{formulated_problem}) into a single-objective optimization problem. After the objectives are normalized, the weighted summation is carried out, and the immediate reward is defined as
    \begin{align}
    \label{REWARD_ORI}
        \begin{split}
            r_v^t(\mathbf{o}_v^t,a_v^t)=-(w_T \cdot \widetilde{T}_{v,m}^{sum}(t)&+w_s \cdot \widetilde{\mathrm{Var}}(\mathbf{S}^o(t))\\
            &+w_d \cdot \xi_v^t ),  
        \end{split}
    \end{align}
    where $w_T$, $w_s$, and $w_d$ represent the weights of optimization objectives, $\widetilde{T}_{v,m}^{sum}(t)$ represents the latency $T_{v,m}^{sum}(t)$ after Z-score normalization in the time dimension, and $\widetilde{\mathrm{Var}}(\mathbf{S}^o(t))$ represents the workload variance after Z-score normalization in the time dimension.

    \item \textbf{Reward Shaping:} Reward shaping is an effective method to incorporate human knowledge into DRL agents \cite{NEURIPS2020_b7109157}. To speed up the model training and convergence, we use the reward shaping method to guide DRL agents in identifying optimal VT migration routes. Specifically, the nonlinearity of variance makes it difficult for agents to improve the variance reward. Therefore, we replace the variance term $\widetilde{\mathrm{Var}}(\mathbf{S}^o(t))$ in Eq. (\ref{REWARD_ORI}) with $\widetilde{S}_m(t)$, and Eq. (\ref{REWARD_ORI}) can be reformulated as
    \begin{align}
        \begin{split}
            r_v^t(\mathbf{o}_v^t,a_v^t)=-(w_T \cdot \widetilde{T}_{v,m}^{sum}(t)&+w_s \cdot \widetilde{S}_m(t)\\
            &+w_d \cdot \xi_v^t ),
        \end{split}
    \end{align}
    where $\widetilde{S}_m(t)$ denotes the workload of edge server $m$ after Z-score normalization.

    \begin{figure}[t]
        \centering
        \includegraphics[width=0.45\textwidth]{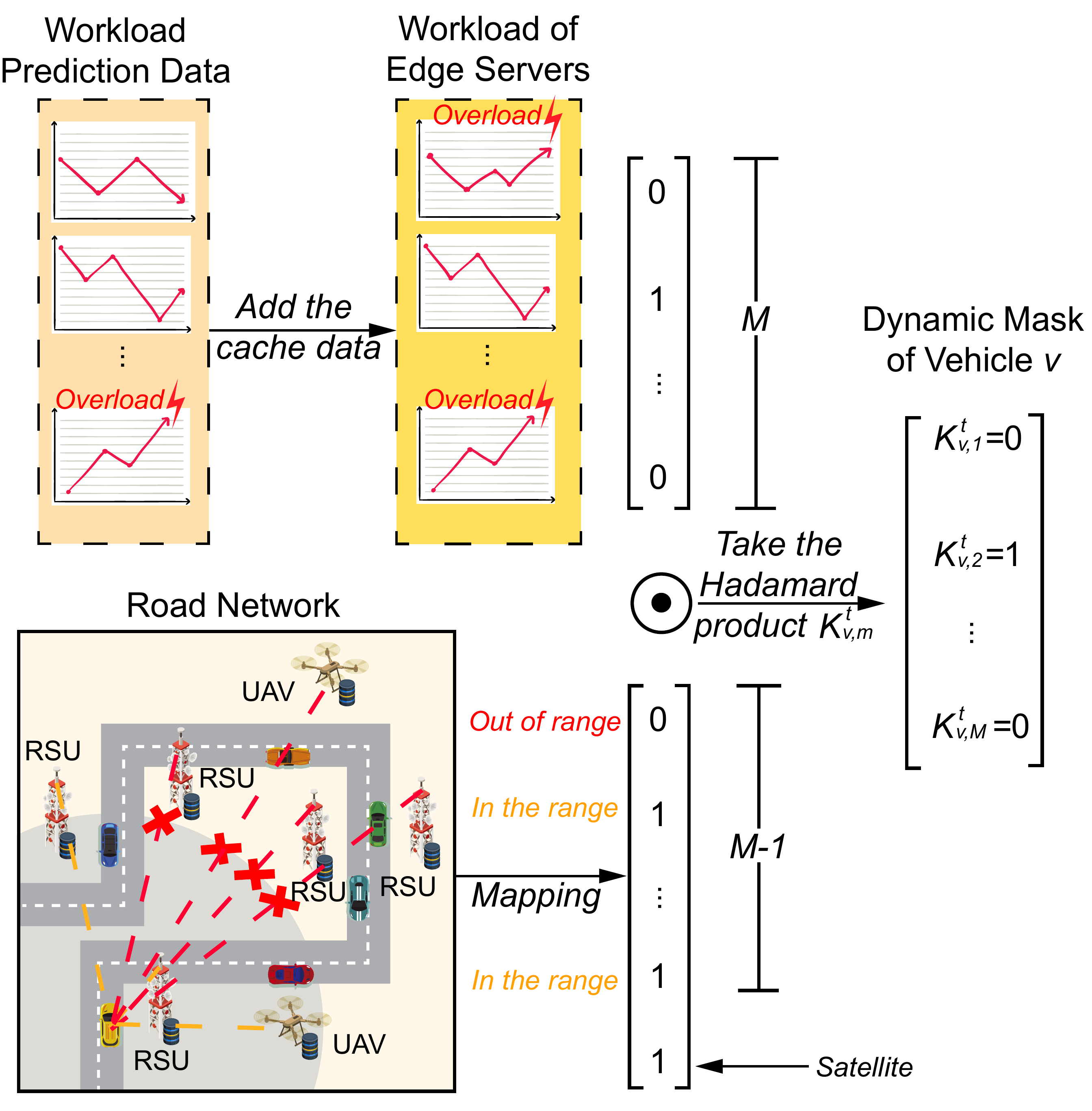}
        \caption{The structure of the dynamic mask module.}
        \label{DM}
    \end{figure}
    
   \item \textbf{Dynamic Mask Module:}
   When VTs of vehicle $v$ migrated to edge server $m$ based on action $a_{v,m}^t$ lead to $S_m'(t)>L_m^{max}$ or $D_{v,m}(t)>D_m^{max}$ in time slot $t$, the action $a_{v,m}^t$ are infeasible. We define the infeasible action set as $\mathcal{M}_{v,t}^{infeasible}$. To prevent DRL agents from choosing an infeasible action, we utilize the dynamic mask module to set the probability of infeasible actions to $0$ \cite{10026882, 10818642}, which accelerates agents to identify optimal migration routes. We define the mask list $\mathbf{K}_v^t =[K_{v,1}^t, \ldots,K_{v,m}^t,\ldots,K_{v,M}^t]$, where $K_{v,m}^t$ is given by
    \begin{equation}
        K_{v,m}^t=
        \begin{cases} 
            0, &  m \in \mathcal{M}_{v,t}^{infeasible},  \\
            1, &  \text{otherwise}.
        \end{cases}
    \end{equation}
    Thus, the sampled action $a_v^t$ for vehicle $v$ is rewritten as
    \begin{equation}
        a_v^t=\underset{m}{\mathrm{argmax}} (\mathbf{P}_v^t \mathbf{K}_v^t).
    \end{equation}

    Figure \ref{DM} shows the complete calculation process of the dynamic mask module. After mapping the output of the LSTM-based Transformer prediction model to workload data, the dynamic mask module will add the task data stored in the cache to obtain the actual workload of edge server $m$ in time slot $t$. This is because the unfinished task in the last time slot $t-1$ will be stored in the cache. Finally, the dynamic mask module will determine if the edge server is about to be overloaded based on the size of the VT migration task. In addition, to ensure that VTs are not migrated to edge servers too far away, the dynamic mask module performs the same mask operation on edge servers beyond the maximum service range $D_m^{max}$.
\end{itemize}

\begin{algorithm}[t]
\DontPrintSemicolon
\SetAlgoLined
\caption{DM-MAPPO for VT Migration Routing}
Initialize actor network parameters $\theta_v$, $\theta_v^{old}$, and critic network parameters $w_v$;\label{AI_Contract}\\
Initialize environment $Env$ and replay buffer $Buf$;\\
\For{$\text{Episode} = 1,2,\ldots,End$}
    {
        Input historical trajectories to the LSTM-based Transformer model.\\
        Store predicted workloads into $Buf$.\\
        \For{$v=1,2,\ldots,V$}
            {
                \For{$t=0,1,\ldots,t_v^{end}$}
                    {
                        Get the observation $\mathbf{\widetilde{o}}_v^t$ from $Env$.\\
                        Add the predicted workload into $\mathbf{\widetilde{o}}_v^t$ to get the observation $\mathbf{o}_v^t$.\\
                        Each agent gets the action from $\pi_{\theta^{old}}$ according to $\mathbf{o}_v^t$.\\
                        Mask the infeasible action.\\
                        Get the reward $r_v^t$ and the next observation $\mathbf{\widetilde{o}}_v^{t+1}$ from $Env$.\\ 
                        Store $\{\mathbf{o}_v^t,\mathbf{o}_v^{t+1},a_v^t,r_v^t \}$ into $Buf$.\\
                    }
            Get the reward list $\left\{ r_v^t \right\}_{t=0}^{t_v^{end}}$ and the next observation list $\left\{ \mathbf{o}_v^{t+1} \right\}_{t=0}^{t_v^{end}}$ from $Buf$.\\
            Calculate the next value list $\left\{ Q_{w_v}(\mathbf{o}_v^{t+1},a_v^{t+1}) \right\}_{t=0}^{t_v^{end}}$ as ${target}$.\\
            Calculate the value list $\left\{ Q_{w_v}(\mathbf{o}_v^t,a_v^t) \right\}_{t=0}^{t_v^{end}}$ as ${value}$.\\
            Calculate the critic network loss as $J_v(w) = \mathrm{MSE}({target},{value}) $ using Eq. (\ref{Jvw}).\\
            Update $w_v$ using $J_v(w)$ according to Eq. (\ref{wv}).\\
            Calculate ${\delta} \leftarrow {target}-{value}$.\\
            Calculate $A_v^{\theta} \leftarrow \sum_{t=0}^{t_v^{end}}{\delta[t] \cdot \gamma^{t-1}}$.\\
            Get $\pi_{\theta^{old}} $ from the actor network.\\
            \For{${epoch}=1,2,\ldots,end$}
            {
                Get $\pi_{\theta}$ from the actor network.\\
                Calculate ${ratio}$ according to Eq. (\ref{ratio}).\\
                Calculate $J_v^{clip}$ according to Eq. (\ref{Jvclip}).\\
                Calculate the actor network loss $J_v(\theta_v^{old})$ according to Eq. (\ref{Jv}).\\
                Update $\theta_v^{old} \leftarrow \theta_v$.\\
                Update $\theta_v$ using $J_v(\theta_v)$ according to Eq. (\ref{thetav}).\\
            }
        }
    }
$\bf{return}$ the optimal migration decision sequence $\left[a_v^0,a_v^1,\ldots,a_v^t,\ldots,a_v^t\right]_{v=1}^{V}$.
\end{algorithm}

\subsection{Algorithm Design}
PPO is an on-policy algorithm based on the classic Actor-Critic (AC) framework, which significantly reduces the sampling time required for environment interaction by repeatedly training on batch-collected data \cite{DBLPjournals/corr/SchulmanWDRK17}. Furthermore, by constraining the difference between the new policy and the old policy, the PPO algorithm ensures stable policy updates, thereby enhancing learning efficiency and accelerating convergence \cite{10416899}. In the AC framework, $Q_{w_v}(\mathbf{o}_v^t,a_v^t)$ is used to estimate the reward expectation in the future after the action $a_v^t$ is executed based on
observation $\mathbf{o}_v^t$ , and $Q_{w_v}(\mathbf{o}_v^t,a_v^t)$ is
\begin{align}
    \begin{split}
    \ Q_{w_v}(\mathbf{o}_v^t,a_v^t)&=\gamma Q_{w_v}(\mathbf{o}_v^{t+1},a_v^{t+1})+r_v^t(\mathbf{o}_v^t,a_v^t) \\
    &\approx \sum_{t'=t}^{t_v^{end}}{\gamma^{t'-t}r_v^{t'}}(\mathbf{o}_v^t,a_v^t),  
    \end{split}
\end{align}
where $\gamma$ represents the discount factor.

The AC framework leverages the critic network to fit the reward distribution to judge the action quality generated by the actor network. Thus, the loss function of the critic network is denoted as
\begin{align}
    \begin{split}
    \label{Jvw}
     &\ J_v(w) = \\
     &\frac{1}{t_v^{end}}\sum_{t=0}^{t_v^{end}}{\left[ -Q_{w_v}\left(\mathbf{o}_v^t,a_v^t \right) + \sum_{t'=t}^{t_v^{end}}{\gamma^{t'-t}r_v^{t'} \left(\mathbf{o}_v^{t'},a_v^{t'} \right)} \right]^2},
    \end{split}
\end{align}
and the parameters of the critic network $w_v$ are updated as
\begin{equation}
\label{wv}
    w_v=w_v+\alpha_w \nabla_w J_v(w).
\end{equation}
where $\alpha_w$ and $\nabla_w J_v(w)$ represent the learning rate and gradient of the critic network, respectively.

PPO algorithms limit the update rate to ensure that the deviation is within a limited range and define the proportion of the change in the action probability distribution as
\begin{equation}
\label{ratio}
    {ratio}=\frac{\pi_{\theta}(a_v^t | \mathbf{o}_v^t)}{\pi_{\theta^{old}}(a_v^t | \mathbf{o}_v^t)}.
\end{equation}

To eliminate the influence of positive and negative rewards on the probability of action exploration, the advantage function on the output of the critic network can be calculated as
\begin{equation}
\label{Av}
    A_v^{\theta} \left( \mathbf{o}_v^t,a_v^t \right)=Q_{w_v}\left( \mathbf{o}_v^t,a_v^t \right)-b_{w_v}\left( \mathbf{o}_v^t \right),
\end{equation}
where the baseline $b_{w_v}\left( \mathbf{o}_v^t \right)$ is the expected reward for all actions taken at the observation $\mathbf{o}_v^t$, which is calculated as
\begin{equation}
\label{bwv}
    b_{w_v}\left( \mathbf{o}_v^t \right)={\mathbb{E}_{a_v^t \in \mathcal{A}_v^t}\left[Q_{w_v}\left(\mathbf{o}_v^t,a_v^t \right)|\mathbf{o}_v^t \right]},
\end{equation}
where $\mathcal{A}_v^t$ represents the set of all actions of vehicle $v$ in time slot $t$, and $\mathbb{E}_{a_v^t \in \mathcal{A}_v^t}[\cdot]$ denotes the expected value.

In addition, the updated range is limited to $(1-\epsilon,1+\epsilon)$, which is calculated as
\begin{equation}
\label{Jvclip}
    J_v^{clip}=\mathrm{clip}({ratio},1-\epsilon,1+\epsilon) \cdot A_v^{\theta} \left( \mathbf{o}_v^t,a_v^t \right).
\end{equation}

Therefore, the loss function of the actor network can be denoted as \cite{DBLPjournals/corr/SchulmanWDRK17}
\begin{equation}
\label{Jv}
    J_v(\theta_v)=\mathbb{E}_t \left\{{\min \left[{ratio} \cdot A_v^{\theta} \left( \mathbf{o}_v^t,a_v^t \right),J_v^{clip} \right]} \right\},
\end{equation}
and the parameters of the actor network $\theta_v$ are updated as
\begin{equation}
\label{thetav}
    \theta_v=\theta_v+\alpha_{\theta} \nabla_{\theta} J_v(\theta_v),
\end{equation}
where $\alpha_{\theta}$ represents the learning rate of the actor network.

Algorithm \ref{AI_Contract} illustrates the pseudocode of the DM-MAPPO algorithm, whose computational complexity is $\mathcal{O}\left(End \times V \times t_v^{end} \times (M + end)\right)$, where $End$ represents the number of complete interactions of all agents with the environment, and $end$ represents the training epochs of each actor network.

\section{Simulation Results}\label{numericalresult}
\subsection{Parameter Settings}

{We first present the parameter settings for the experiments in the VT migration routing scenario. The scenario is sampled every $30\:\rm{s}$. Since all UAVs are designed to reach the next RSU for replenishment with the least energy consumption and in the shortest time, the trajectories of all UAVs are fixed, and the UAVs are set to patrol among RSUs at a speed of $3.7\:\rm{m/s}$\cite{8119562}. For the network model, the mean of additive Gaussian white noise $\sigma$ is set to $0$, with a variance of $10$ \cite{10185562}. The vehicle trajectories are obtained from the dataset\footnote{Vehicle trajectory dataset: \url{https://www.microsoft.com/en-us/research/publication/t-drive-trajectory-data-sample/}}, and the traffic flow data are sourced from the dataset\footnote{ California traffic flow dataset: \url{https://pems.dot.ca.gov/}}. The main parameters used in the experiments are summarized in Table \ref{parameter} \cite{9511221,10185562,9558857,8712145}.}

\begin{table}[t]
	\renewcommand{\arraystretch}{1.2}
        \captionsetup{font = normal}
	\caption{ Key Parameter Setting in the Paper. } \centering \label{parameter}
	\begin{tabular}{m{5.4cm}<{\raggedright}|m{2.0cm}<{\centering}}	 	
		\hline		
		\textbf{Hyperparameters} & \textbf{Setting}\\	
		\hline
		The number of vehicles $(C)$ &  $5$\\	
		\hline
		The number of UAVs $(U)$ &  $10$\\
		\hline
		The number of RSUs ($N$)   & $68\sim 84$  \\
		\hline
		  Migration task size ($S_v^{mig}$) &  $60\:\rm{MB}$\\	
		\hline		
		CPU operating frequency of UAVs and RSUs ($C_m$)&  $32 \times 10^9\:\rm{Hz}$\\
		\hline
            CPU operating frequency of the satellite ($C_s$)&  $80 \times 10^9\:\rm{Hz}$ \\
		\hline
            Maximum cache capacity of the edge server ($L_m^{max}$) &  $300 \:\rm{MB}$\\
		\hline
            Vehicle transmitting power ($p_v$) &  $1000\:\rm{W}$\\
		\hline
            Uplink bandwidth ($B_m^{up}$) &  $300\:\rm{MHz}$\\
		\hline
            Downlink bandwidth ($B_m^{down}$) &  $300\:\rm{MHz}$\\
		\hline
            Maximum service distance of RSU ($D_m^{max},m\in \mathcal{N}_v^t$) &  $1200\:\rm{m}$\\
		\hline
            Maximum service distance of UAV ($D_m^{max},m\in \mathcal{U}_v^t$) &  $1500\:\rm{m}$\\
		\hline
            UAV cruising altitude ($h_u$) &  $50\:\rm{m}$\\
		\hline
            The number of CPU cycles required per unit data ($e_v$) &  $1000\:\rm{cycles/bit}$\\
		\hline
            {The weight for the latency objective ($w_T$)} &  {$2$}\\
		\hline
            {The weight for the workload variance objective ($w_s$)} &  {$1$}\\
		\hline
            {The weight for the packet loss rate objective ($w_d$)} &  {$1$}\\
		\hline
	\end{tabular}
\end{table}

To demonstrate the effectiveness of the proposed LSTM-based Transformer model, we compare it against four baseline models: LSTM, GRU, CNN-GRU, and CNN-LSTM. In both CNN-GRU and CNN-LSTM, the input data is initially processed using a one-dimensional convolution with a kernel size of $\mathrm{1}$, followed by prediction using GRU or LSTM, respectively \cite{10.1145/3377713.3377722,app13074644}. For the LSTM-based Transformer model, considering the factors of training time and prediction accuracy, we set the window of historical data input as $T=6$ to predict the traffic flow data in the next time slot $t+1$. In particular, for the LSTM prediction layer, the hidden unit in the LSTM module is set to $d_h=32$. For the Transformer prediction layer, the number of attention heads is set to $H=4$, and the dimension of query, key, and value vectors is set to $d_k=64$. 

{To demonstrate the effectiveness of the proposed DM-MAPPO algorithm, we compare it with several baseline algorithms, including Multi-Agent Soft Actor-Critic (MASAC), Greedy, and Random algorithms\cite{10944431}. The Greedy algorithm selects the edge server geographically closest to the vehicle for VT migration. To enhance exploration and avoid infeasible actions, a $\mathrm{10\%}$ probability is introduced for randomly selecting a server. The Random algorithm performs VT migration by randomly selecting the next edge server without any optimization or learned strategy. To maintain environmental stability during the training of the DRL algorithm, the LSTM-based Transformer model should be trained before the DRL algorithm. This ensures that the workload of each edge server observed by the DRL algorithm remains consistent.
}

{
All experiments presented in this paper are conducted on a hardware platform equipped with an Intel Core i5 CPU and an MX450 GPU.
}

\subsection{{Performance of LSTM-based Transformer Model on Workload Prediction}}
\begin{figure}[t]
\centering
\includegraphics[width=0.48\textwidth]{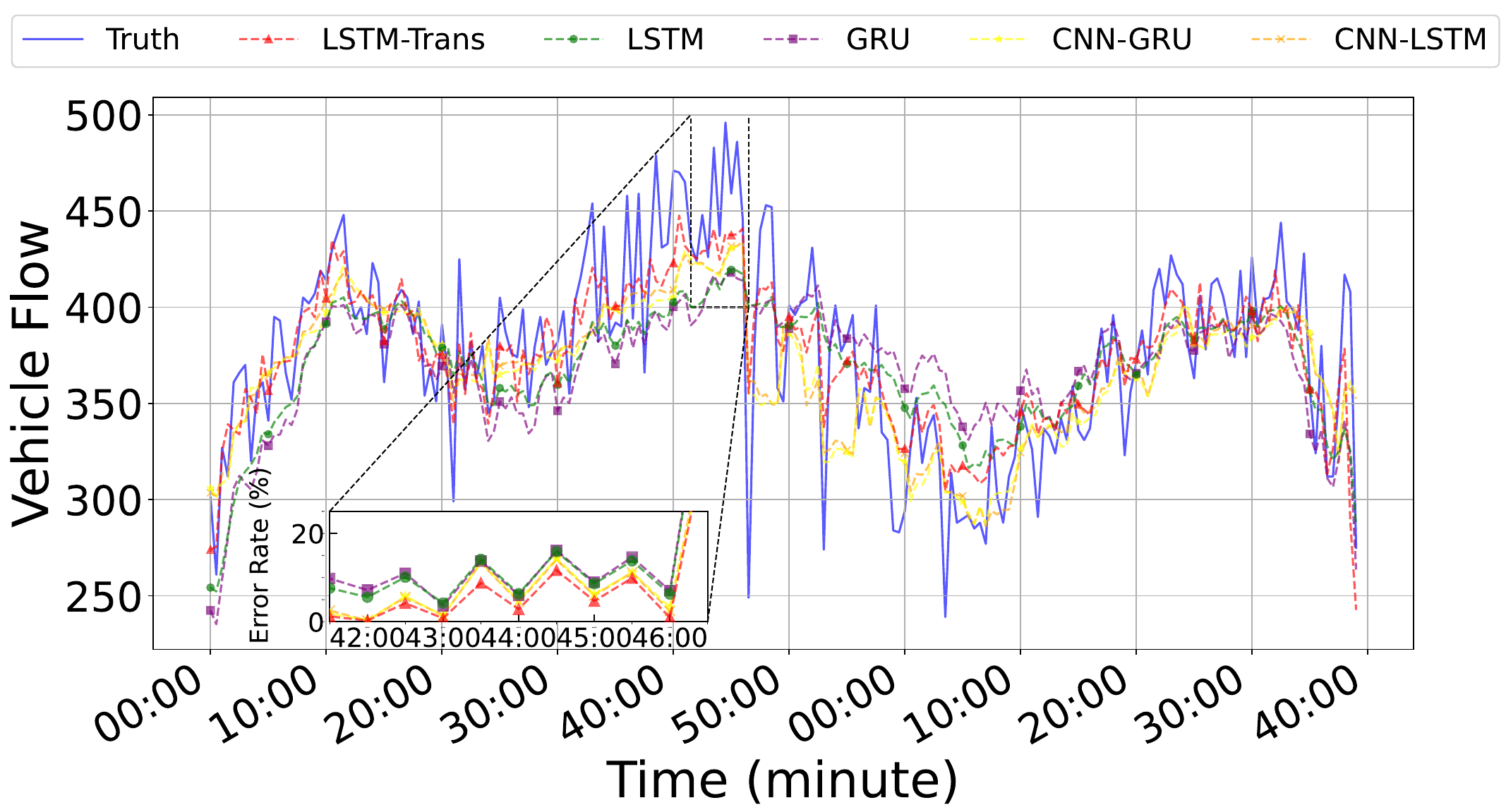}
\caption{The workload of edge servers predicted by different prediction algorithms over time.}
\label{pre_point}
\end{figure}

In Fig. \ref{pre_point}, we show the predicted RSU workload curve generated by five different prediction algorithms. Among them, the proposed LSTM-based Transformer model yields a prediction curve that most closely aligns with the actual RSU workload. Figure \ref{pre_point} also shows the error rate of each prediction model in the high-frequency part of the traffic flow data, where the LSTM-based Transformer prediction error is always the smallest, outperforming the other models. As shown in Table \ref{The performance of different methods}, the proposed model achieves the lowest Root Mean Square Error (RMSE) and Mean Absolute Error (MAE), indicating the smallest deviation between the predicted and actual values. Specifically, the proposed LSTM-based Transformer model achieves a prediction error rate of $\mathrm{6.814\%}$, highlighting its superior prediction performance. Moreover, the determination $R^2$ of the model approaches $1$, demonstrating strong explanatory power and the ability to capture most of the variance in the workload data. {The LSTM-based Transformer model outperforms all baselines across all four evaluation metrics. This is because the LSTM module possesses the powerful capability of processing time-series data. By first leveraging LSTM to capture temporal dependencies and reduce high-frequency noise, followed by the self-attention mechanism of Transformer for fine-grained prediction refinement, the model delivers highly accurate results, and its efficient training time of only $4$ minutes on the hardware platform demonstrates its strong potential for real-world applications.}

\begin{table}[t]
\renewcommand{\arraystretch}{1.6}
 \centering
 \tiny
\captionsetup{font = normal}
 \caption{The Performance of Various Prediction Algorithms.}
 \setlength{\tabcolsep}{3pt}
 {\fontsize{9}{9}\selectfont
  \begin{tabular}{cccccc}
    \toprule[1.5pt]
    \hline
    {Metrics} & \textbf{Ours} & \textbf{CNN-GRU} & \textbf{LSTM} & \textbf{GRU} & \textbf{CNN-LSTM} \\\midrule
    {RMSE} & \textbf{0.671} & 0.812 & 0.877 & 0.916 & 0.917 \\ 
    {MAE}  & \textbf{20.417} & 27.802 & 25.330 & 29.466 & 29.461 \\ 
    {Error Rate} & \textbf{6.814\%} & 9.451\%  & 8.741\% & 9.843\% & 9.796\%  \\
    ${R}^\textbf{2}$ & \textbf{0.925} & 0.865 & 0.884 & 0.851 & 0.855 \\\hline
    \bottomrule[1.5pt]
 \end{tabular}
 \selectfont
 }
 \label{The performance of different methods}
\end{table}

\subsection{Performance of DM-MAPPO on VT Migration Routing}
\subsubsection{Performance of LSTM-based Transformer models for VT migration route planning}

\begin{figure}[t]
    \centering
    \includegraphics[width=0.48\textwidth]{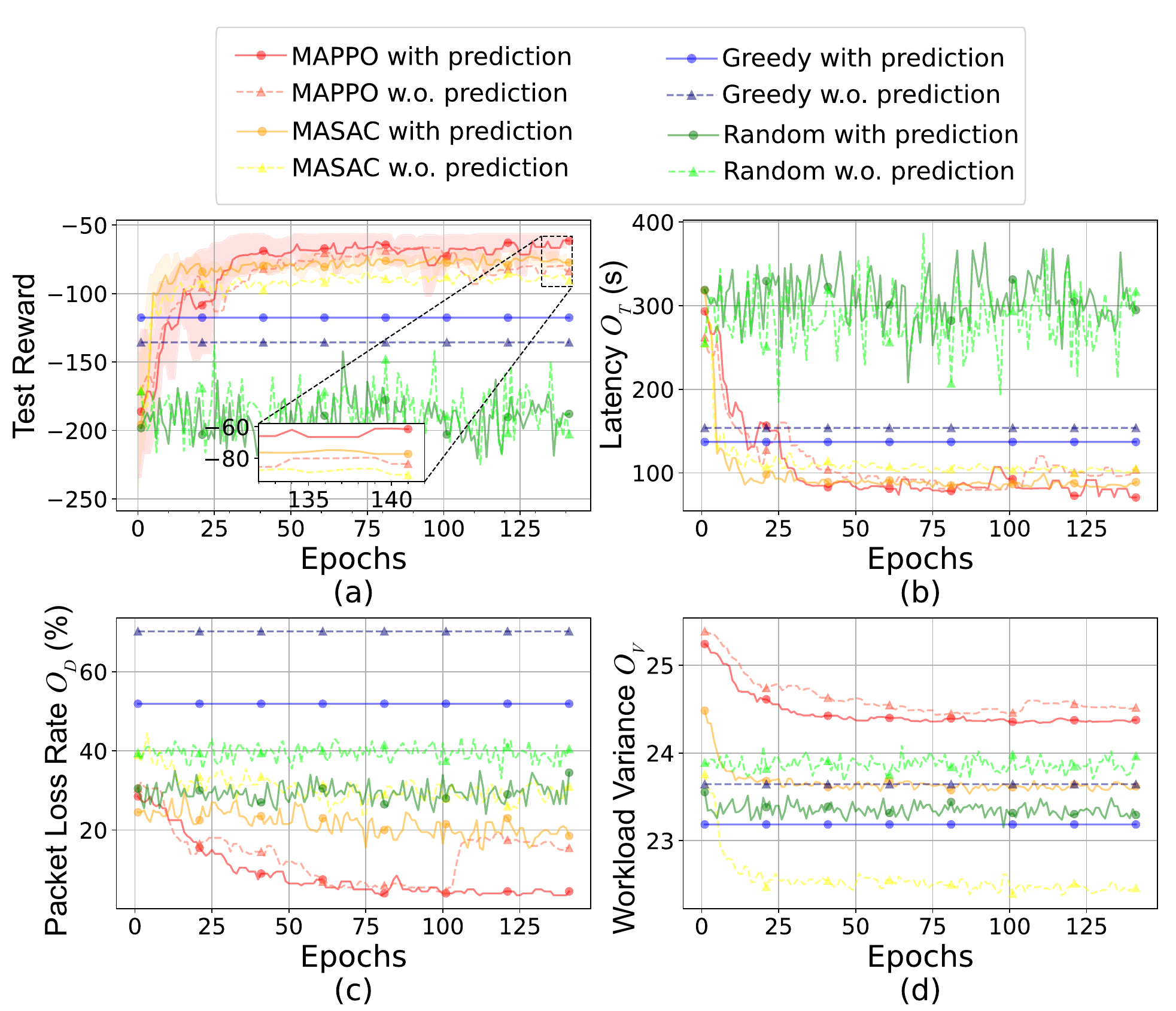}
    \caption{Performance comparison of test reward curves for various DRL algorithms under the dynamic mask module.}
    \label{preandnopre}
\end{figure}

\begin{figure}[t]
    \centering
    \includegraphics[width=0.48\textwidth]{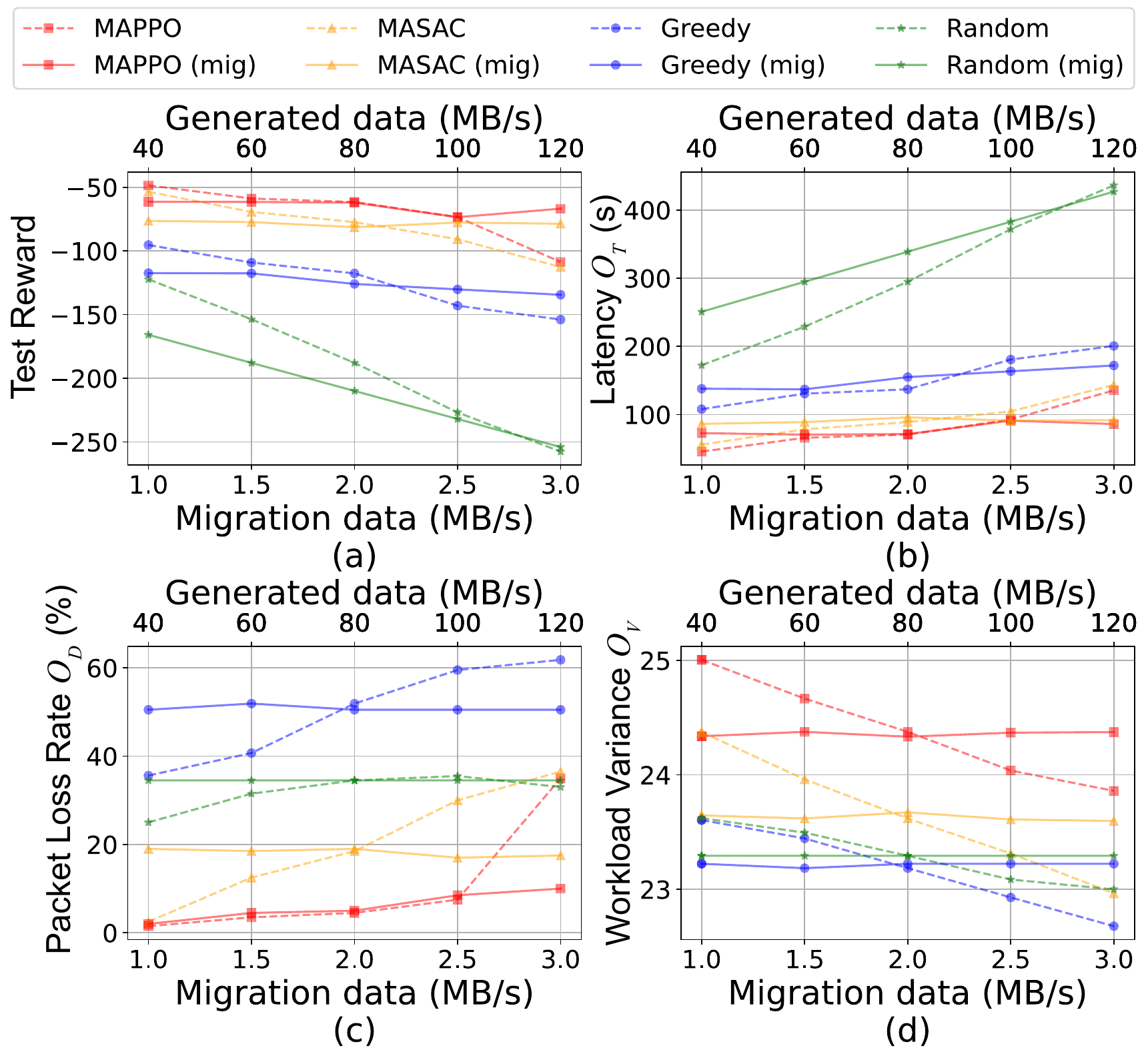}
    \caption{Performance comparison of different DRL algorithms in multiple evaluation metrics, including test rewards, latency, packet loss rates, and variances of workloads.}
    \label{MM}
\end{figure}

As shown in Fig. \ref{preandnopre} (a), we comprehensively evaluate the performance of the proposed DM-MAPPO algorithm with and without the LSTM-based Transformer prediction model. For ease of expression, we call them \textit{MAPPO with prediction} and \textit{MAPPO w.o. prediction}, respectively. The test reward achieved by \textit{MAPPO with prediction} is $\mathrm{35.99\%}$ higher than that of \textit{MAPPO w.o. prediction}. Similarly, the test reward achieved by \textit{MASAC with prediction} is $\mathrm{17.08\%}$ higher than that of \textit{MASAC w.o. prediction}. The above results indicate that the integration of the LSTM-based Transformer model significantly reduces the likelihood of infeasible actions taken by the DM-MAPPO algorithm. In Fig. \ref{preandnopre} (b), the total latency $T_{v,m}^{sum}(t)$ achieved by \textit{MAPPO with prediction} is $\mathrm{70.4729\:\rm{s}}$, which corresponds to an average transmission latency of $8.42\:\rm{ms}$ per vehicle per data transmission. This is lower than the $\mathrm{12.40\:\rm{ms}}$ latency obtained using \textit{MAPPO w.o. prediction}. Furthermore, Fig. \ref{preandnopre} (c) shows that \textit{MAPPO with prediction} achieves the lowest packet loss rate of $\mathrm{4.494\%}$, ensuring a high-quality immersive experience. In terms of workload balancing, Fig. \ref{preandnopre} (d) shows that \textit{MAPPO with prediction} reduces the workload variance of the edge server cluster to $24.38$, and \textit{MAPPO w.o. prediction} to $24.53$, which proves that the LSTM-based Transformer model can balance the workload. This shows that our proposed \textit{MAPPO with prediction} effectively reduces transmission delay and packet loss while improving workload distribution, enhancing the overall immersive experience for users.

\begin{figure}[t]
    \centering
    \includegraphics[width=0.48\textwidth]{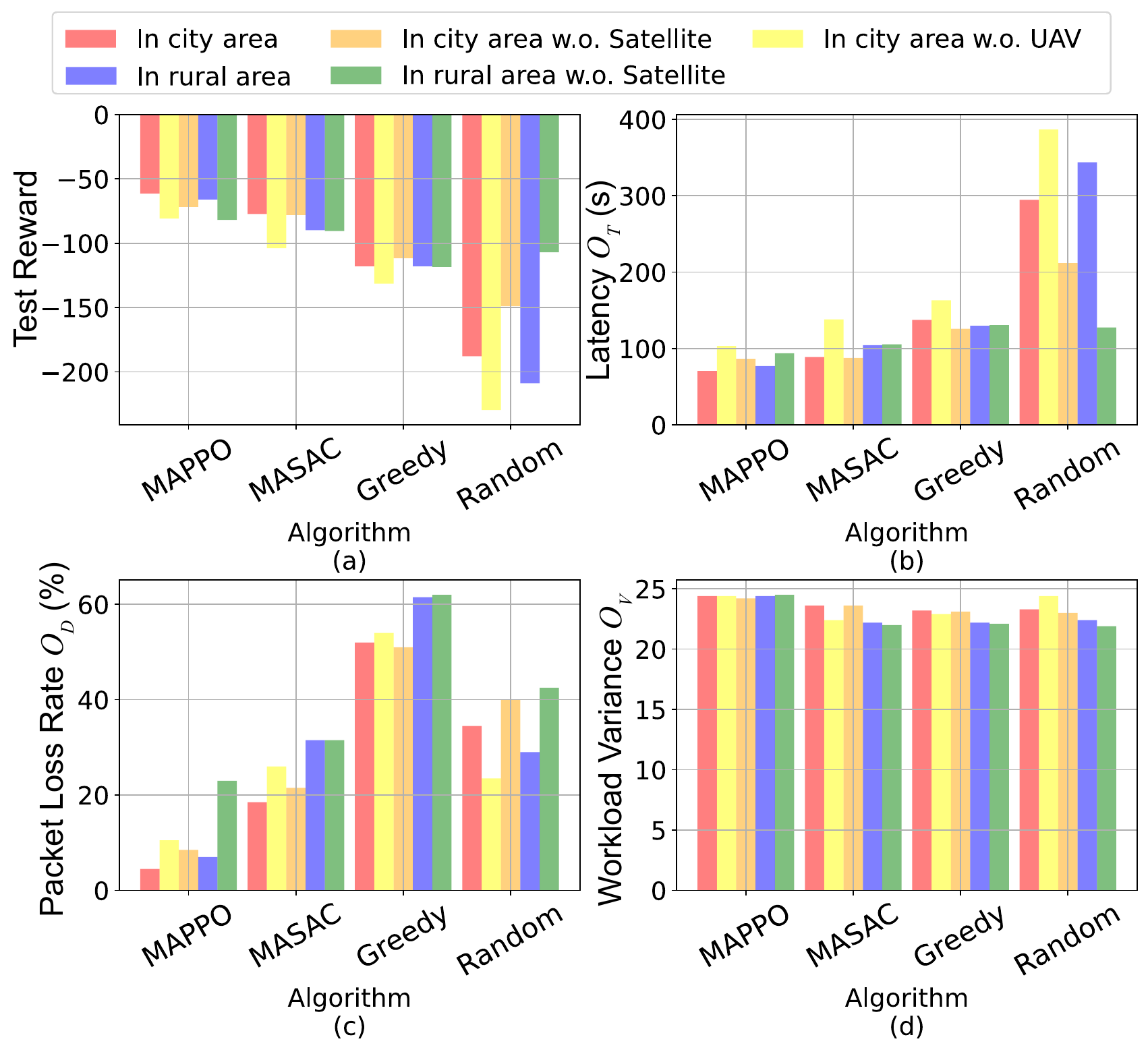}
    \caption{The ablation of different framework elements.}
    \label{CITY}
\end{figure}

\subsubsection{Performance of DM-MAPPO on various VT migration tasks}
As shown in Fig. \ref{MM}, the test reward of each algorithm decreases as the migration data increases. This is due to the inevitable increase in computation and communication times caused by larger migration data packets. For migration data rates of $\mathrm{40\:\rm{MB/s}}$, $\mathrm{60\:\rm{MB/s}}$, $\mathrm{80\:\rm{MB/s}}$, $\mathrm{100\:\rm{MB/s}}$, and $\mathrm{120\:\rm{MB/s}}$, our proposed DM-MAPPO algorithm consistently achieves the lowest latency and packet loss rate, as shown in Fig. \ref{MM} (b) and Fig. \ref{MM} (c). Specifically, when the migration data rate is at or below $\mathrm{100\:\rm{MB/s}}$, the packet loss rate of DM-MAPPO remains below $\mathrm{10\%}$, indicating that the proposed DM-MAPPO algorithm ensures stable and reliable data transmission during VT migration. Furthermore, to assess the impact of varying computational requirements, we conduct experiments with tasks that generate different volumes of data. In Fig. \ref{MM} (d), the workload variance decreases with the increase of generated data. This is because each vehicle can cause a greater impact on the edge server network. Thus, when serving vehicles, the workload of edge servers with low workloads will be increased, and the workload variance will be reduced. For generated data rates of $\mathrm{1.0\:\rm{MB/s}}$, $\mathrm{1.5\:\rm{MB/s}}$, $\mathrm{2.0\:\rm{MB/s}}$, $\mathrm{2.5\:\rm{MB/s}}$, and $\mathrm{3.0\:\rm{MB/s}}$, our proposed DM-MAPPO can achieve the lowest latency and packet loss rate, which demonstrates that DM-MAPPO can reasonably migrate VTs under different computational task requirements to minimize both total latency and packet loss.

\subsubsection{Ablation study on framework elements}

We conduct an ablation study by independently removing UAVs and the satellite from the proposed framework to evaluate their respective contributions to VT migration. As shown in Fig. \ref{CITY}, in the city area without UAVs, the test reward of the four algorithms decreases by an average of $\mathrm{6.90\%}$, while the average latency and packet loss rate increase by $\mathrm{37.88\%}$ and $\mathrm{29.78\%}$, respectively. The reason is that when RSUs are overloaded, VTs can be migrated to UAVs, avoiding a complete interruption of vehicular services, which indicates that UAVs can alleviate the uneven distribution of RSUs and the problem of insufficient computing resources during peak hours in the city area. In addition, in the city area, DM-MAPPO can further reduce the latency and packet loss rate of VT migration with satellites, while the performance of all baseline algorithms declines with satellite due to the high cost of satellite communication, resulting in high latency and packet loss rates. Compared with the city area, the performance of each algorithm decreases in the rural area environment. Fortunately, in rural areas with satellites, the latency and packet loss rates of VT migration are lower than in rural areas without satellites, suggesting that satellites can compensate for the lack of RSU support for communication services in remote areas.

\subsubsection{Ablation study on the effectiveness of the dynamic mask module}

\begin{figure}[t]
    \centering
    \includegraphics[width=0.48\textwidth]{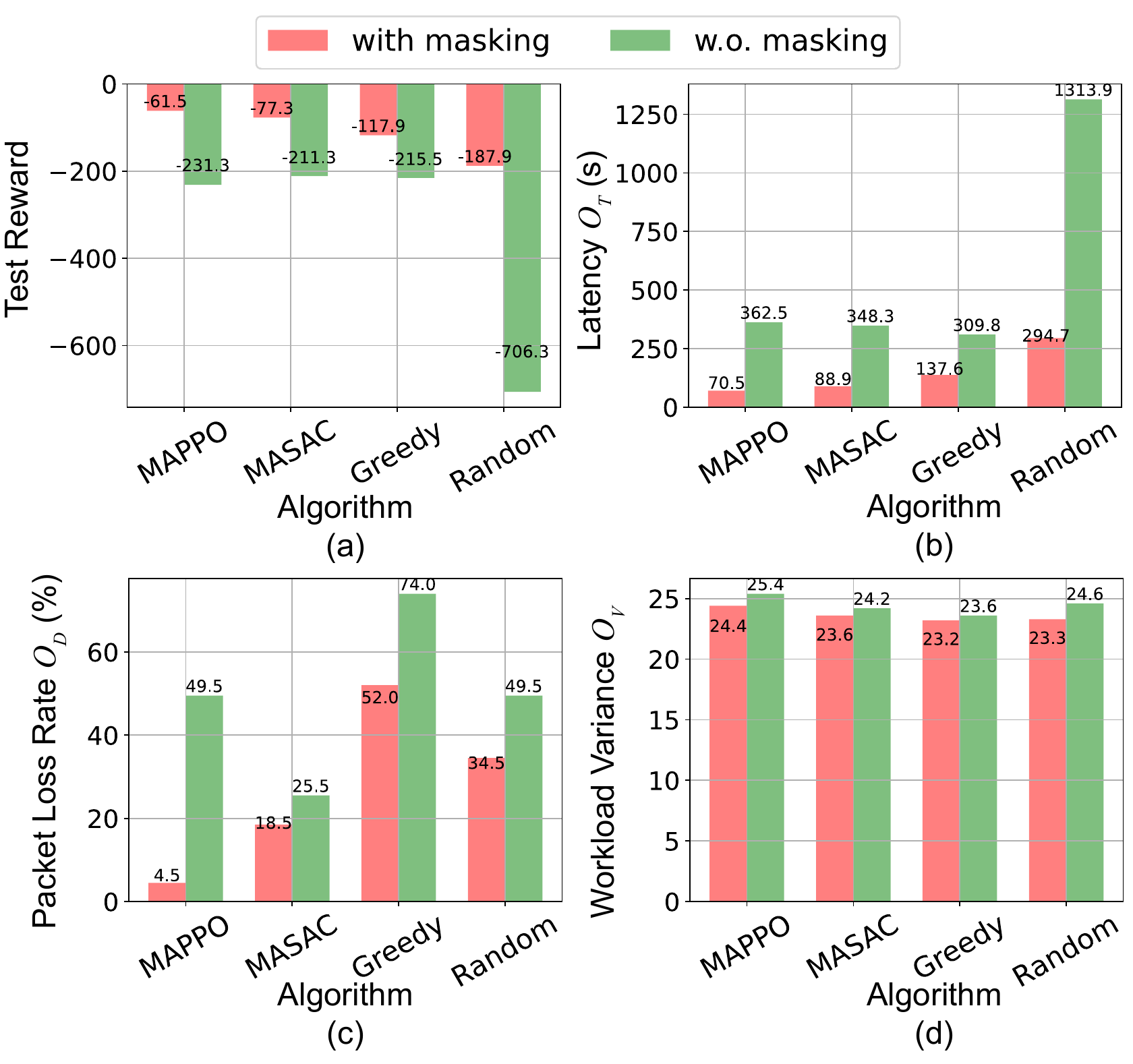}
    \caption{Comparison of performance gains brought by the dynamic mask module.}
    \label{MASK}
\end{figure}

\begin{figure}[t]
    \centering
    \includegraphics[width=0.48\textwidth]{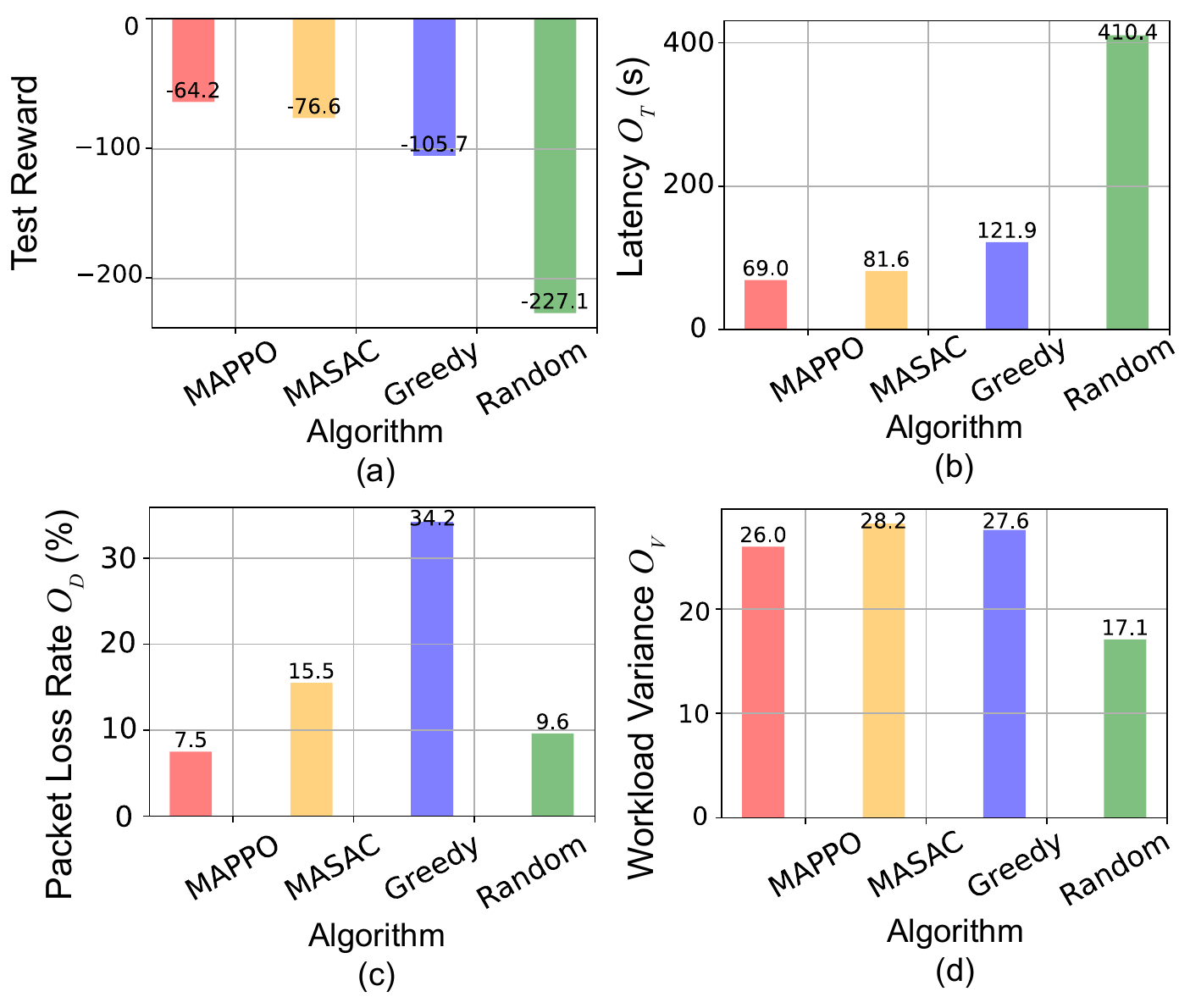}
    \caption{{Algorithm performance comparison in scenarios with varying numbers of UAVs.}}
    \label{CHANGE}
\end{figure}

\begin{figure}[t]
    \centering
    \includegraphics[width=0.48\textwidth]{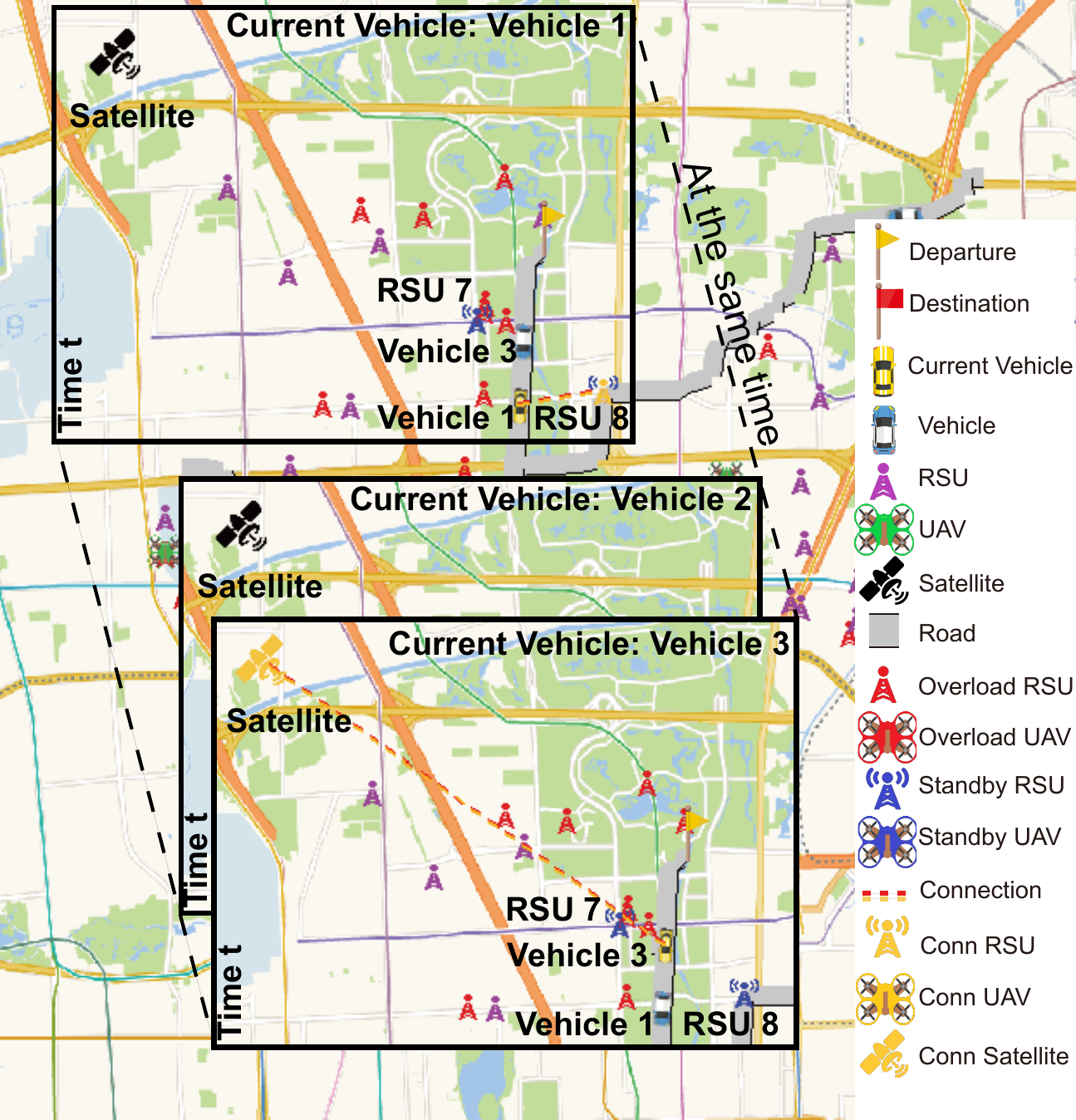}
    \caption{Cooperative behavior exhibited by multiple PPO agents when planning multiple VT migration routes in the same area.}
    \label{Avoid}
\end{figure}

To assess the performance improvement brought by the dynamic mask module for each algorithm, we conduct an ablation study by removing this module from four algorithms, allowing VTs to be migrated to any edge server without constraints. As shown in Fig. \ref{MASK}, algorithms with the masking module significantly outperform those without the masking module in terms of latency and packet loss rates, with an average latency reduction of $\mathrm{72.05\%}$ and an average packet loss rate reduction of $\mathrm{44.60\%}$. Particularly, the dynamic mask module brings a $\mathrm{73.41\%}$ performance improvement to MAPPO, enabling it to surpass MASAC. Moreover, the dynamic mask module can reduce workload variance, balancing the workload of RSUs in the region.

\begin{figure*}[t]
    \centering
    \includegraphics[width=0.98\textwidth]{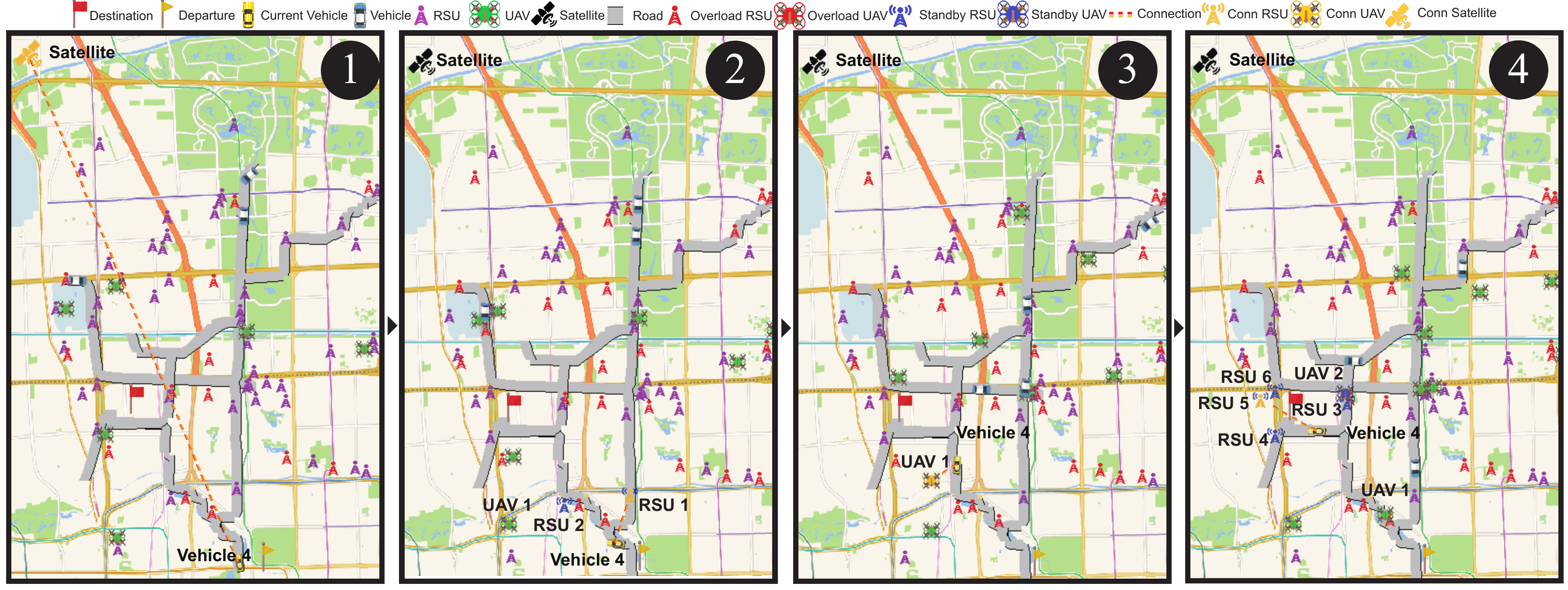}
    \caption{Single VT migration process in real scenarios.}
    \label{Migration}
\end{figure*}

\subsubsection{{Comparison of algorithm robustness under different numbers of edge servers}}

{
In practical scenarios, the total number of edge servers in an area is not fixed. This is mainly because UAVs may exit the region before vehicles reach their destinations, leading to dynamic variability in the number of available UAVs. Fortunately, with the introduction of a masking mechanism, the output of the action network can be expanded based on the currently known number of edge servers, with unused output entries being set to zero. The mask remains active until new UAVs enter the area. This approach provides a flexible interface for connecting intermittently arriving or departing UAVs to the DRL network, ensuring consistent integration under dynamic conditions.
}

{
The experimental setup is configured as follows: $5$ vehicles depart at time slot $0$, while the remaining $2$ vehicles depart at different time slots, simulating dynamically entering vehicles. Among the edge servers, $5$ UAVs appear from the edge of the scenario at various times and begin providing VT services. In addition, $6$ UAVs move toward different edges of the area at different times and exit the region.
As shown in Fig. \ref{CHANGE}, DM-MAPPO still maintains the best performance under these experimental conditions, achieving the lowest latency and packet loss rate. This demonstrates that DM-MAPPO possesses strong robustness, enabling it to effectively adapt to highly dynamic environments, and thus making it well-suited for real-world applications.
}

\subsubsection{Visual analysis of multi-agent cooperative behavior}

{To demonstrate the cooperation among agents in DM-MAPPO during the VT migration process, we develop a visualization platform to display the migration of each VT intuitively. Typically, the number of vehicles exceeds that of RSUs. However, the edge servers providing services to these five vehicles are also capable of serving other vehicles. For a simple and clear presentation, we have exclusively displayed the behavioral trajectories of these $5$ vehicles controlled by the DM-MAPPO algorithm. As shown in Fig.\ref{Avoid}, in time slot $t$, both vehicle $1$ and vehicle $3$ are traveling from north to south within the same region. As they approach the coverage boundary of RSU $7$, they need to migrate their VTs to edge servers with better channel quality, thus ensuring the immersion of VT services. For the vehicle $1$, RSU $8$, and the satellite are identified as feasible edge server targets for VT migration. Given the higher communication cost associated with satellite links, vehicle $1$ opts to migrate its VTs to RSU $8$. Since the agents in DM-MAPPO share information such as workload information, vehicle $3$ migrating its VTs to RSU $8$ will overload RSU $8$, which may cause packet loss and affect the VT services of other vehicles. Therefore, for vehicle $3$, migrating its VTs to the satellite can maximize the overall benefit for all vehicles, which shows that our proposed DM-MAPPO can solve cooperative tasks well.}


\subsubsection{Visual analysis of the VT migration process}
In Fig. \ref{Migration}, we present the complete VT migration trajectory of vehicle $4$ within the target region. At moment $1$, there are no RSUs or UAVs in the standby state. Therefore, the optimal option for vehicle $4$ is to migrate its VTs to the satellite. At moment $2$, vehicle $4$ enters the service range of RSU $1$ and RSU $2$. To avoid the high costs of satellite services, the vehicle $4$ migrates its VTs from the satellite to the RSU $1$. At moment $3$, RSU $2$ is overloaded, and vehicle $4$ is out of the service range of RSU $1$ but enters the service range of UAV $1$. Since the communication costs for UAVs are significantly lower than those for the satellite, the optimal option for vehicle $4$ is to migrate its VTs to UAV $1$. At moment $4$, vehicle $4$ is out of the service range of UAV $1$ and migrates its VTs to RSU $5$. This is because RSU $5$ is the closest RSU to the destination, and vehicle $4$ does not need to migrate VTs again, which can minimize total latency. This indicates that the actions taken by our proposed DM-MAPPO are highly interpretable, the migration process is clear, and the planned VT migration path is reasonable and reliable.

\section{Conclusion}\label{conclusion}
{In this paper, we have addressed the challenges of optimizing multi-objective VT migration routing in 6G-enabled IoV networks. Specifically, we have proposed a workload prediction-based VT migration framework that leverages a heterogeneous edge server architecture comprising RSUs, UAVs, and the satellite. This architecture is particularly effective in mitigating the problem of low resource utilization in sparsely distributed regions. To enable accurate long-term workload prediction, we have proposed the LSTM-based Transformer model that integrates the LSTM module into the Transformer model. Furthermore, we have proposed the DM-MAPPO algorithm based on the dynamic mask module to facilitate efficient VT migration route planning in complex environments. The DM-MAPPO algorithm can greatly reduce the probability of infeasible actions and comprehensively optimize key performance metrics such as latency, packet loss rate, and regional workload balance. Simulation results have demonstrated that the LSTM-based Transformer model can reduce the error rate to $\mathrm{6.814\%}$, achieving accurate traffic flow prediction, and the performance of the DM-MAPPO algorithm has been improved by $\mathrm{73.41\%}$ compared with MAPPO. The proposed framework operates under the idealistic assumptions of fully trustworthy edge servers and simplified channel models. However, these assumptions may lead to risks such as the exposure of VT migration data and vehicle trajectories, potentially hindering the framework's practical application. Therefore, for future work, we will explore the channel model with line-of-sight and non-line-of-sight natures and incorporate privacy-preserving mechanisms into our framework\cite{10608129,8077766}.} 

\bibliographystyle{IEEEtran}
\bibliography{ref}

@ARTICLE{10416899,
  author={Kang, Jiawen and Zhong, Yue and Xu, Minrui and Nie, Jiangtian and Wen, Jinbo and Du, Hongyang and Ye, Dongdong and Huang, Xumin and Niyato, Dusit and Xie, Shengli},
  journal={IEEE Internet of Things Journal}, 
  title={Tiny Multiagent {DRL} for Twins Migration in {UAV} Metaverses: A Multileader Multifollower Stackelberg Game Approach}, 
  year={2024},
  volume={11},
  number={12},
  pages={21021-21036},
  doi={10.1109/JIOT.2024.3360183}}

@ARTICLE{10818642,
  author={Wen, Jinbo and Kang, Jiawen and Niyato, Dusit and Zhang, Yang and Mao, Shiwen},
  journal={IEEE Transactions on Industrial Cyber-Physical Systems}, 
  title={{Sustainable Diffusion-Based Incentive Mechanism for Generative {AI}-Driven Digital Twins in Industrial Cyber-Physical Systems}}, 
  year={2025},
  volume={3},
  number={},
  pages={139-149},
  doi={10.1109/TICPS.2024.3524483}}

@inproceedings{NEURIPS2020_b7109157,
 author = {Hu, Yujing and Wang, Weixun and Jia, Hangtian and Wang, Yixiang and Chen, Yingfeng and Hao, Jianye and Wu, Feng and Fan, Changjie},
 booktitle = {Advances in Neural Information Processing Systems},
 editor = {H. Larochelle and M. Ranzato and R. Hadsell and M.F. Balcan and H. Lin},
 pages = {15931--15941},
 publisher = {Curran Associates, Inc.},
 title = {{Learning to Utilize Shaping Rewards: A New Approach of Reward Shaping}},
 volume = {33},
 year = {2020}
}

@INPROCEEDINGS{10271832,
  author={Wen, Jinbo and Kang, Jiawen and Xiong, Zehui and Zhang, Yang and Du, Hongyang and Jiao, Yutao and Niyato, Dusit},
  booktitle={2023 IEEE International Conference on Metaverse Computing, Networking and Applications (MetaCom)}, 
  title={{Task Freshness-Aware Incentive Mechanism for Vehicle Twin Migration in Vehicular Metaverses}}, 
  year={2023},
  volume={},
  number={},
  pages={481-487},
  doi={10.1109/MetaCom57706.2023.00089}}

@article{DBLPjournals/corr/SchulmanWDRK17,
  author       = {John Schulman and
                  Filip Wolski and
                  Prafulla Dhariwal and
                  Alec Radford and
                  Oleg Klimov},
  title        = {{Proximal Policy Optimization Algorithms}},
  journal      = {CoRR},
  volume       = {abs/1707.06347},
  year         = {2017},
  eprinttype    = {arXiv},
  eprint       = {1707.06347},
  bibsource    = {dblp computer science bibliography, https://dblp.org  }
}

@article{YADAV2024122333,
title = {{{NOA-LSTM}: An Efficient {LSTM} Cell Architecture for Time Series Forecasting}},
journal = {Expert Systems with Applications},
volume = {238},
pages = {122333},
year = {2024},
issn = {0957-4174},
doi = {https://doi.org/10.1016/j.eswa.2023.122333  },
author = {Hemant Yadav and Amit Thakkar}}

@ARTICLE{10526476,
  author={Mao, Bomin and Zhou, Xueming and Liu, Jiajia and Kato, Nei},
  journal={IEEE Transactions on Vehicular Technology}, 
  title={{On a Cooperative Deep Reinforcement Learning-Based Multi-Objective Routing Strategy for Diversified 6G Metaverse Services}}, 
  year={2024},
  volume={73},
  number={9},
  pages={14092-14096},
  doi={10.1109/TVT.2024.3397707}}

@inproceedings{zhong2024no,
  title={{No {P}rior {M}ask: {E}liminate {R}edundant {A}ction for {D}eep {R}einforcement {L}earning}},
  author={Zhong, Dianyu and Yang, Yiqin and Zhao, Qianchuan},
  booktitle={Proceedings of the AAAI Conference on Artificial Intelligence},
  volume={38},
  number={15},
  pages={17078--17086},
  year={2024}
}

@article{vaswani2017attention,
  title={{Attention is all you need}},
  author={Vaswani, A},
  journal={Advances in Neural Information Processing Systems},
  year={2017}
}

@ARTICLE{9300168,
  author={Ren, Pei and Qiao, Xiuquan and Huang, Yakun and Liu, Ling and Pu, Calton and Dustdar, Schahram and Chen, Junliang},
  journal={IEEE Transactions on Cloud Computing}, 
  title={{Edge {AR} {X5}: {A}n {E}dge-{A}ssisted {M}ulti-{U}ser {C}ollaborative {F}ramework for {M}obile {W}eb {A}ugmented {R}eality in 5G and {B}eyond}}, 
  year={2022},
  volume={10},
  number={4},
  pages={2521-2537},
  doi={10.1109/TCC.2020.3046128}}

@ARTICLE{6773024,
  author={Shannon, C. E.},
  journal={The Bell System Technical Journal}, 
  title={{A {M}athematical {T}heory of {C}ommunication}}, 
  year={1948},
  volume={27},
  number={3},
  pages={379-423},
  doi={10.1002/j.1538-7305.1948.tb01338.x}}

@Article{app13074644,
AUTHOR = {Song, Hyunsun and Choi, Hyunjun},
TITLE = {{Forecasting {S}tock {M}arket {I}ndices {U}sing the {R}ecurrent {N}eural {N}etwork {B}ased {H}ybrid {M}odels: {CNN}-{LSTM}, {GRU}-{CNN}, and {E}nsemble {M}odels}},
JOURNAL = {Applied Sciences},
VOLUME = {13},
YEAR = {2023},
NUMBER = {7},
ARTICLE-NUMBER = {4644},
ISSN = {2076-3417},
DOI = {10.3390/app13074644}
}

@inproceedings{10.1145/3377713.3377722,
author = {Yamak, Peter T. and Yujian, Li and Gadosey, Pius K.},
title = {{A {C}omparison {B}etween {ARIMA}, {LSTM}, and {GRU} for {T}ime {S}eries {F}orecasting}},
year = {2020},
isbn = {9781450372619},
publisher = {Association for Computing Machinery},
address = {New York, NY, USA},
doi = {10.1145/3377713.3377722},
booktitle = {Proceedings of the 2019 2nd International Conference on Algorithms, Computing and Artificial Intelligence},
pages = {49–55},
numpages = {7},
location = {Sanya, China},
series = {ACAI '19}
}

@ARTICLE{9511221,
  author={Dinc, Ergin and Vondra, Michal and Cavdar, Cicek},
  journal={IEEE Transactions on Vehicular Technology}, 
  title={{Total {C}ost of {O}wnership {O}ptimization for {D}irect {A}ir-to-{G}round {C}ommunication {N}etworks}}, 
  year={2021},
  volume={70},
  number={10},
  pages={10157-10172},
  doi={10.1109/TVT.2021.3103634}}

@ARTICLE{10185562,
  author={Chen, Junlong and Kang, Jiawen and Xu, Minrui and Xiong, Zehui and Niyato, Dusit and Chen, Chuan and Jamalipour, Abbas and Xie, Shengli},
  journal={IEEE Internet of Things Journal}, 
  title={{Multiagent {D}eep {R}einforcement {L}earning for {D}ynamic {A}vatar {M}igration in {AIoT}-{E}nabled {V}ehicular {M}etaverses {W}ith {T}rajectory {P}rediction}}, 
  year={2024},
  volume={11},
  number={1},
  pages={70-83},
  doi={10.1109/JIOT.2023.3296075}}

@ARTICLE{9944868,
  author={Xu, Minrui and Ng, Wei Chong and Lim, Wei Yang Bryan and Kang, Jiawen and Xiong, Zehui and Niyato, Dusit and Yang, Qiang and Shen, Xuemin and Miao, Chunyan},
  journal={IEEE Communications Surveys \& Tutorials}, 
  title={{A {F}ull {D}ive {I}nto {R}ealizing the {E}dge-{E}nabled {M}etaverse: {V}isions, {E}nabling {T}echnologies, and {C}hallenges}}, 
  year={2023},
  volume={25},
  number={1},
  pages={656-700},
  doi={10.1109/COMST.2022.3221119}}

@ARTICLE{8712145,
  author={Wang, Jun and Feng, Daquan and Zhang, Shengli and Tang, Jianhua and Quek, Tony Q. S.},
  journal={IEEE Access}, 
  title={{Computation Offloading for Mobile Edge Computing Enabled Vehicular Networks}}, 
  year={2019},
  volume={7},
  number={},
  pages={62624-62632},
  doi={10.1109/ACCESS.2019.2915959}}

@ARTICLE{9558857,
  author={Masaracchia, Antonino and Li, Yijiu and Nguyen, Khoi Khac and Yin, Cheng and Khosravirad, Saeed R. and Costa, Daniel Benevides Da and Duong, Trung Q.},
  journal={IEEE Access}, 
  title={{UAV-Enabled Ultra-Reliable Low-Latency Communications for {6G}: A Comprehensive Survey}}, 
  year={2021},
  volume={9},
  number={},
  pages={137338-137352},
  doi={10.1109/ACCESS.2021.3117902}}

@article{GUO2024237,
title = {{{Survey} on {Digital} {Twins} for {Internet} of {Vehicles}: {Fundamentals}, {Challenges}, and {Opportunities}}},
journal = {Digital Communications and Networks},
volume = {10},
number = {2},
pages = {237-247},
year = {2024},
issn = {2352-8648},
doi = {<url id="cvht1kp8bjvsl7d97gu0" type="url" status="parsed" title="Survey on digital twins for Internet of Vehicles: Fundamentals, challenges, and opportunities" wc="1607">https://doi.org/10.1016/j.dcan.2022.05.023</url> },
author = {Jiajie Guo and Muhammad Bilal and Yuying Qiu and Cheng Qian and Xiaolong Xu and Kim-Kwang {Raymond Choo}}
}

@incollection{cook2023complexity,
  title={{The} {Complexity} of {Theorem-Proving} {Procedures}},
  author={Cook, Stephen A},
  booktitle={Logic, automata, and computational complexity: The works of Stephen A. Cook},
  pages={143--152},
  year={2023}
}

@ARTICLE{10638123,
  author={Wen, Jinbo and Nie, Jiangtian and Zhong, Yue and Yi, Changyan and Li, Xiaohuan and Jin, Jiangming and Zhang, Yang and Niyato, Dusit},
  journal={IEEE Internet of Things Journal}, 
  title={{Diffusion-Model-Based} {Incentive} {Mechanism} {With} {Prospect} {Theory} for {Edge} {AIGC} {Services} in {6G IoT}}, 
  year={2024},
  volume={11},
  number={21},
  pages={34187-34201},
  doi={10.1109/JIOT.2024.3445171}}

@ARTICLE{8869705,
  author={Saad, Walid and Bennis, Mehdi and Chen, Mingzhe},
  journal={IEEE Network}, 
  title={{A} {Vision} of {6G} {Wireless} {Systems}: {Applications}, {Trends}, {Technologies}, and {Open} {Research} {Problems}}, 
  year={2020},
  volume={34},
  number={3},
  pages={134-142},
  doi={10.1109/MNET.001.1900287}}

@ARTICLE{9628162,
  author={Guo, Hongzhi and Li, Jingyi and Liu, Jiajia and Tian, Na and Kato, Nei},
  journal={IEEE Communications Surveys \& Tutorials}, 
  title={{A} {Survey} on {Space-Air-Ground-Sea} {Integrated} {Network} {Security} in {6G}}, 
  year={2022},
  volume={24},
  number={1},
  pages={53-87},
  doi={10.1109/COMST.2021.3131332}}

@ARTICLE{9880528,
  author={Wang, Yuntao and Su, Zhou and Zhang, Ning and Xing, Rui and Liu, Dongxiao and Luan, Tom H. and Shen, Xuemin},
  journal={IEEE Communications Surveys \& Tutorials}, 
  title={{A} {Survey} on {Metaverse}: {Fundamentals}, {Security}, and {Privacy}}, 
  year={2023},
  volume={25},
  number={1},
  pages={319-352},
  doi={10.1109/COMST.2022.3202047}}

@ARTICLE{9410247,
  author={Chen, Long and Hu, Bin and Guan, Zhi-Hong and Zhao, Lian and Shen, Xuemin},
  journal={IEEE Transactions on Neural Networks and Learning Systems}, 
  title={{Multiagent Meta-Reinforcement Learning for Adaptive Multipath Routing Optimization}}, 
  year={2022},
  volume={33},
  number={10},
  pages={5374-5386},
  doi={10.1109/TNNLS.2021.3070584}}

@ARTICLE{10233845,
  author={Khan, Anwar and Fouda, Mostafa M. and Do, Dinh-Thuan and Almaleh, Abdulaziz and Rahman, Atiq Ur},
  journal={IEEE Access}, 
  title={{Short-Term} {Traffic} {Prediction} {Using} {Deep} {Learning} {Long} {Short-Term} {Memory}: {Taxonomy}, {Applications}, {Challenges}, and {Future} {Trends}}, 
  year={2023},
  volume={11},
  number={},
  pages={94371-94391},
  doi={10.1109/ACCESS.2023.3309601}}

@ARTICLE{9745481,
  author={Dai, Yueyue and Zhang, Yan},
  journal={Journal of Communications and Information Networks}, 
  title={{Adaptive} {Digital} {Twin} for {Vehicular} {Edge} {Computing} and {Networks}}, 
  year={2022},
  volume={7},
  number={1},
  pages={48-59},
  doi={10.23919/JCIN.2022.9745481}}

@article{GULMEZ2023120346,
title = {{Stock {Price} {Prediction} with {Optimized} {Deep} {LSTM} {Network} with {Artificial} {Rabbits} {Optimization} {Algorithm}}},
journal = {Expert {Systems} with {Applications}},
volume = {227},
pages = {120346},
year = {2023},
issn = {0957-4174},
doi = {<url id="cvhtb547fff8jn4tch8g" type="url" status="parsed" title="Stock price prediction with optimized deep LSTM network with artificial rabbits optimization algorithm" wc="2326">https://doi.org/10.1016/j.eswa.2023.120346</url> },
author = {Burak Gülmez}
}

@article{Zhou_Zhang_Peng_Zhang_Li_Xiong_Zhang_2021, 
title={{Informer: Beyond {Efficient} {Transformer} for {Long} {Sequence} {Time-Series} {Forecasting}}}, 
volume={35},
DOI={10.1609/aaai.v35i12.17325},
number={12}, 
journal={Proceedings of the AAAI Conference on Artificial Intelligence},
author={Zhou, Haoyi and Zhang, Shanghang and Peng, Jieqi and Zhang, Shuai and Li, Jianxin and Xiong, Hui and Zhang, Wancai},
year={2021},
month={May},
pages={11106-11115} }

@inproceedings{liu2022pyraformer,
  author = {Liu, S. and Yu, H. and Liao, C. and Li, J. and Lin, W. and Liu, A. X. and Dustdar, S.},
  title = {{Pyraformer: Low-Complexity Pyramidal Attention for Long-Range Time Series Modeling and Forecasting}},
  booktitle = {The Tenth International Conference on Learning Representations (ICLR 2022)},
  year = {2022},
}

@ARTICLE{9726783,
  author={Liu, Tong and Tang, Lun and Wang, Weili and He, Xiaoqiang and Chen, Qianbin and Zeng, Xiaoping and Jiang, Haitao},
  journal={IEEE Internet of Things Journal}, 
  title={{Resource {Allocation} in {DT-assisted} {Internet} of {Vehicles} via {Edge} {Intelligent} {Cooperation}}}, 
  year={2022},
  volume={9},
  number={18},
  pages={17608-17626},
  doi={10.1109/JIOT.2022.3156100}}

@article{10.1145/3532611,
author = {Li, Fuxian and Feng, Jie and Yan, Huan and Jin, Guangyin and Yang, Fan and Sun, Funing and Jin, Depeng and Li, Yong},
title = {{Dynamic {Graph} {Convolutional} {Recurrent} {Network} for {Traffic} {Prediction}: {Benchmark} and {Solution}}},
year = {2023},
issue_date = {January 2023},
publisher = {Association for Computing Machinery},
address = {New York, NY, USA},
volume = {17},
number = {1},
issn = {1556-4681},
doi = {10.1145/3532611},
journal = {ACM Trans. Knowl. Discov. Data},
month = feb,
articleno = {9},
numpages = {21}
}

@ARTICLE{10026882,
  author={Kang, Hongyue and Chang, Xiaolin and Mišić, Jelena and Mišić, Vojislav B. and Fan, Junchao and Liu, Yating},
  journal={IEEE Internet of Things Journal}, 
  title={{Cooperative {UAV} {Resource} {Allocation} and {Task} {Offloading} in {Hierarchical} {Aerial} {Computing} {Systems}: {A} {MAPPO-Based} {Approach}}}, 
  year={2023},
  volume={10},
  number={12},
  pages={10497-10509},
  doi={10.1109/JIOT.2023.3240173}}

@ARTICLE{10034418,
  author={Zhou, Xuanhong and Bilal, Muhammad and Dou, Ruihan and Rodrigues, Joel J. P. C. and Zhao, Qingzhan and Dai, Jianguo and Xu, Xiaolong},
  journal={IEEE Transactions on Intelligent Transportation Systems}, 
  title={{Edge {Computation} {Offloading} With {Content} {Caching} in 6G-Enabled {IoV}}}, 
  year={2024},
  volume={25},
  number={3},
  pages={2733-2747},
  doi={10.1109/TITS.2023.3239599}}

@ARTICLE{10572232,
  author={Ning, Xiangrui and Zeng, Ming and Hua, Meng and Fei, Zesong},
  journal={IEEE Transactions on Vehicular Technology}, 
  title={{Multiple {Reconfigurable} {Intelligent} {Surfaces} Aided {Vehicular} {Edge} {Computing} {Networks}: A {MAPPO-Based} {Approach}}}, 
  year={2024},
  volume={73},
  number={11},
  pages={17496-17509},
  doi={10.1109/TVT.2024.3419554}}

@article{kang2024hybrid,
  title={{Hybrid-{Generative} {Diffusion} {Models} for {Attack-Oriented} {Twin} {Migration} in {Vehicular} {Metaverses}}},
  author={Kang, Yingkai and Wen, Jinbo and Kang, Jiawen and Zhang, Tao and Du, Hongyang and Niyato, Dusit and Yu, Rong and Xie, Shengli},
  journal={arXiv preprint arXiv:2407.11036},
  year={2024}
}

@ARTICLE{9737450,
  author={Xu, Xiaolong and Jiang, Qinting and Zhang, Peiming and Cao, Xuefei and Khosravi, Mohammad R. and Alex, Linss T. and Qi, Lianyong and Dou, Wanchun},
  journal={IEEE Transactions on Fuzzy Systems}, 
  title={{Game {Theory} for {Distributed} {IoV} {Task} {Offloading} With {Fuzzy} {Neural} {Network} in {Edge} {Computing}}}, 
  year={2022},
  volume={30},
  number={11},
  pages={4593-4604},
  doi={10.1109/TFUZZ.2022.3158000}}

@ARTICLE{10251544,
  author={Ernest, Tan Zheng Hui and Madhukumar, A. S.},
  journal={IEEE Transactions on Mobile Computing}, 
  title={{Computation {Offloading} in {MEC-Enabled} {IoV} {Networks}: {Average} {Energy} {Efficiency} {Analysis} and {Learning-Based} {Maximization}}}, 
  year={2024},
  volume={23},
  number={5},
  pages={6074-6087},
  doi={10.1109/TMC.2023.3315275}}

@article{ADNAN2024100615,
title = {{Fundamental {Design} {Aspects} of {UAV-enabled} {MEC} {Systems}: A {Review} on {Models}, {Challenges}, and {Future} {Opportunities}}},
journal = {Computer Science Review},
volume = {51},
pages = {100615},
year = {2024},
issn = {1574-0137},
doi = {<url id="cvhtov596bk2re65096g" type="url" status="parsed" title="Fundamental design aspects of UAV-enabled MEC systems: A review on models, challenges, and future opportunities" wc="20617">https://doi.org/10.1016/j.cosrev.2023.100615</url> },
author = {Mohd Hirzi Adnan and Zuriati Ahmad Zukarnain and Oluwatosin Ahmed Amodu}
}

@ARTICLE{9806434,
  author={Song, Wei and Rajak, Shaik and Dang, Shuping and Liu, Ruijun and Li, Jun and Chinnadurai, Sunil},
  journal={IEEE Transactions on Intelligent Transportation Systems}, 
  title={{Deep {Learning} {Enabled} {IRS} for {6G} {Intelligent} {Transportation} {Systems}: A {Comprehensive} {Study}}}, 
  year={2023},
  volume={24},
  number={11},
  pages={12973-12990},
  doi={10.1109/TITS.2022.3184314}}

@ARTICLE{10551388,
  author={Yuan, Xiaoming and Zhang, Wenyuan and Yang, Jiayu and Xu, Minrui and Niyato, Dusit and Deng, Qingxu and Li, Changle},
  journal={IEEE Internet of Things Journal}, 
  title={{Efficient {IoV} {Resource} {Management} {Through} {Enhanced} {Clustering}, {Matching}, and {Offloading} in {DT-Enabled} {Edge} {Computing}}}, 
  year={2024},
  volume={11},
  number={18},
  pages={30172-30186},
  doi={10.1109/JIOT.2024.3410176}}

@article{chen2023long,
  title={{Long {Sequence} {Time-series} {Forecasting} with {Deep} {Learning}: A {Survey}}},
  author={Chen, Zonglei and Ma, Minbo and Li, Tianrui and Wang, Hongjun and Li, Chongshou},
  journal={Information Fusion},
  volume={97},
  pages={101819},
  year={2023},
  publisher={Elsevier}
}

@ARTICLE{10288593,
  author={ALMahadin, Ghayth and Aoudni, Yassine and Shabaz, Mohammad and Agrawal, Anurag Vijay and Yasmin, Ghazaala and Alomari, Esraa Saleh and Al-Khafaji, Hamza Mohammed Ridha and Dansana, Debabrata and Maaliw, Renato Racelis},
  journal={IEEE Transactions on Consumer Electronics}, 
  title={{VANET {Network} {Traffic} {Anomaly} {Detection} {Using} {GRU-Based} {Deep} {Learning} {Model}}}, 
  year={2024},
  volume={70},
  number={1},
  pages={4548-4555},
  doi={10.1109/TCE.2023.3326384}}

@article{ZHOU2023103886,
title = {{Expanding the {Prediction} {Capacity} in {Long} {Sequence} {Time-series} {Forecasting}}},
journal = {Artificial Intelligence},
volume = {318},
pages = {103886},
year = {2023},
issn = {0004-3702},
doi = {<url id="cvhtov596bk2re650970" type="url" status="parsed" title="Expanding the prediction capacity in long sequence time-series forecasting" wc="18313">https://doi.org/10.1016/j.artint.2023.103886</url> },
author = {Haoyi Zhou and Jianxin Li and Shanghang Zhang and Shuai Zhang and Mengyi Yan and Hui Xiong}
}

@article{SUN2023105662,
title = {{Lightweight {Bidirectional} {Long} {Short-term} {Memory} {Based} on {Automated} {Model} {Pruning} with {Application} to {Bearing} {Remaining} {Useful} {Life} {Prediction}}},
journal = {Engineering Applications of Artificial Intelligence},
volume = {118},
pages = {105662},
year = {2023},
issn = {0952-1976},
doi = {<url id="cvhtov596bk2re65097g" type="url" status="parsed" title="Lightweight bidirectional long short-term memory based on automated model pruning with application to bearing remaining useful life prediction" wc="16264">https://doi.org/10.1016/j.engappai.2022.105662</url> },
author = {Jiankai Sun and Xin Zhang and Jiaxu Wang}
}

@ARTICLE{10944431,
  author={Xu, Rui and Li, Gaolei and Wu, Jun and Li, Jianhua and Zhao, Yue and Liu, Yuchen and Chen, Mingzhe},
  journal={IEEE Transactions on Wireless Communications}, 
  title={{Toward Covert and Reliable Communication for Anti-Eavesdropping Transmission in V2X Networks}}, 
  year={2025},
  volume={24},
  number={8},
  pages={6429-6442},
  doi={10.1109/TWC.2025.3553132}}

@ARTICLE{10852157,
  author={Yang, Xiao and Li, Gaolei and Wu, Jun and Zhou, Kai and Li, Jianhua and Yang, Wu},
  journal={IEEE Transactions on Consumer Electronics}, 
  title={{Backdoor-Empowered Regulable Privilege Authorization for Edge-Level Graph Learning in 6G Vehicular Networks}}, 
  year={2025},
  volume={71},
  number={2},
  pages={6307-6318},
  doi={10.1109/TCE.2025.3533648}}

@ARTICLE{10608129,
  author={Zhang, Zifan and Fang, Minghong and Chen, Mingzhe and Li, Gaolei and Lin, Xi and Liu, Yuchen},
  journal={IEEE Internet of Things Journal}, 
  title={{Securing Distributed Network Digital Twin Systems Against Model Poisoning Attacks}}, 
  year={2024},
  volume={11},
  number={21},
  pages={34312-34324},
  doi={10.1109/JIOT.2024.3421895}}

@ARTICLE{8077766,
  author={Atzeni, Italo and Arnau, Jesús and Kountouris, Marios},
  journal={IEEE Transactions on Wireless Communications}, 
  title={{Downlink Cellular Network Analysis With LOS/NLOS Propagation and Elevated Base Stations}}, 
  year={2018},
  volume={17},
  number={1},
  pages={142-156},
  doi={10.1109/TWC.2017.2763136}}

@ARTICLE{8119562,
  author={Zhan, Cheng and Zeng, Yong and Zhang, Rui},
  journal={IEEE Wireless Communications Letters}, 
  title={{Energy-Efficient Data Collection in {UAV} {Enabled} Wireless Sensor Network}}, 
  year={2018},
  volume={7},
  number={3},
  pages={328-331},
  doi={10.1109/LWC.2017.2776922}}

@ARTICLE{9861699,
  author={Azari, M. Mahdi and Solanki, Sourabh and Chatzinotas, Symeon and Kodheli, Oltjon and Sallouha, Hazem and Colpaert, Achiel and Mendoza Montoya, Jesus Fabian and Pollin, Sofie and Haqiqatnejad, Alireza and Mostaani, Arsham and Lagunas, Eva and Ottersten, Bjorn},
  journal={IEEE Communications Surveys \& Tutorials}, 
  title={{Evolution of Non-Terrestrial Networks From 5G to 6G: A Survey}}, 
  year={2022},
  volume={24},
  number={4},
  pages={2633-2672},
  doi={10.1109/COMST.2022.3199901}}

@ARTICLE{7932157,
  author={Jeong, Seongah and Simeone, Osvaldo and Kang, Joonhyuk},
  journal={IEEE Transactions on Vehicular Technology}, 
  title={{Mobile Edge Computing via a UAV-Mounted Cloudlet: Optimization of Bit Allocation and Path Planning}}, 
  year={2018},
  volume={67},
  number={3},
  pages={2049-2063},
  doi={10.1109/TVT.2017.2706308}}

\begin{IEEEbiography}
[{\includegraphics[width=1in,height=1.25in,clip,keepaspectratio]{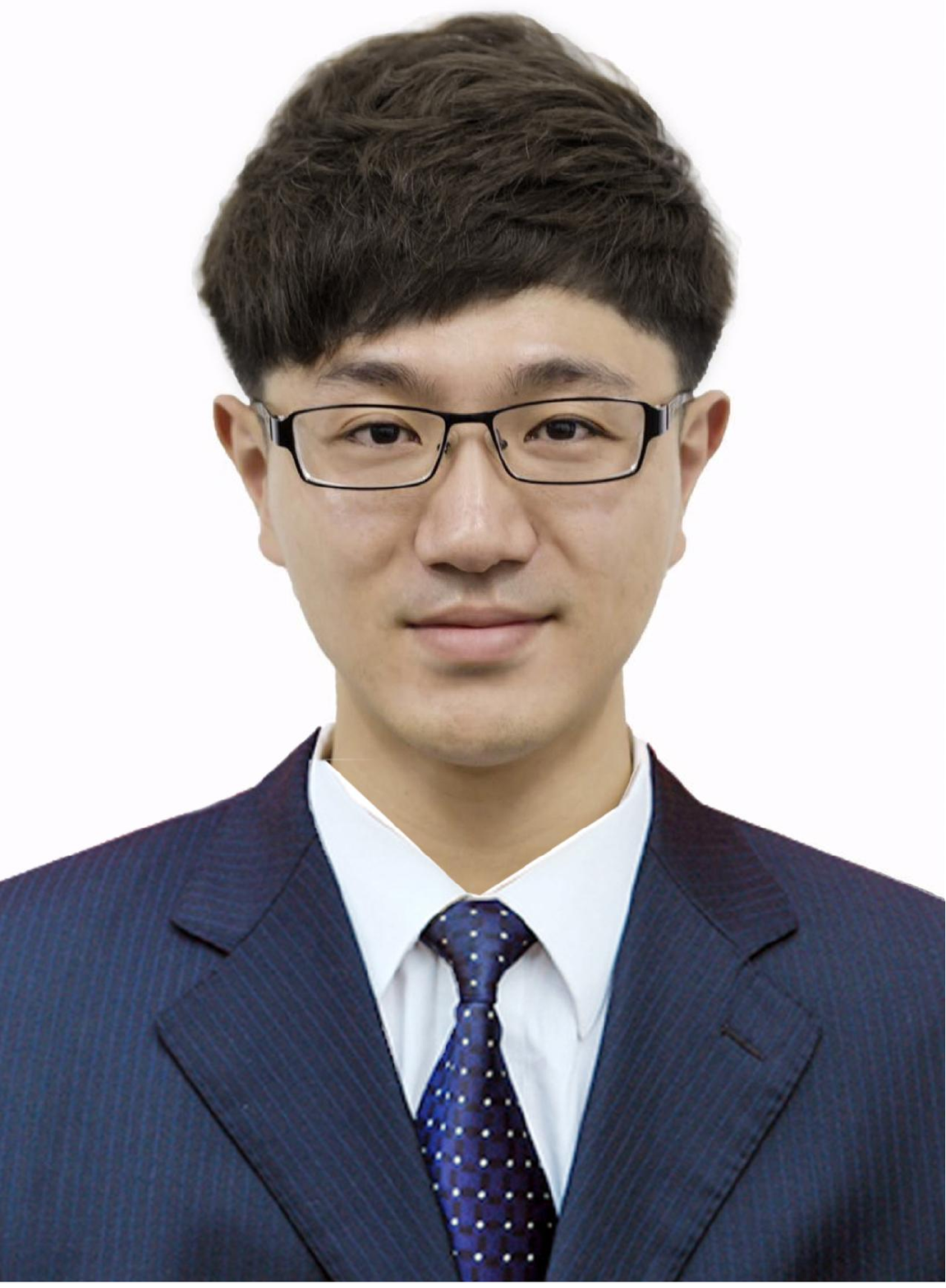}}] 
{Peng Yin} was born in 1989. He received his B.S. and M.S. degrees from Beijing Institute of Technology in 2013 and 2017, respectively. Currently, he is working as an associate research fellow in the Defence Industry Secrecy Examination and Certification Center. His research interests include Electromagnetic cyberspace security, Integrated sensing and communication technology, and Signal detection and recognition technology.
\end{IEEEbiography}

\begin{IEEEbiography}
[{\includegraphics[width=1in,height=1.25in,clip,keepaspectratio]{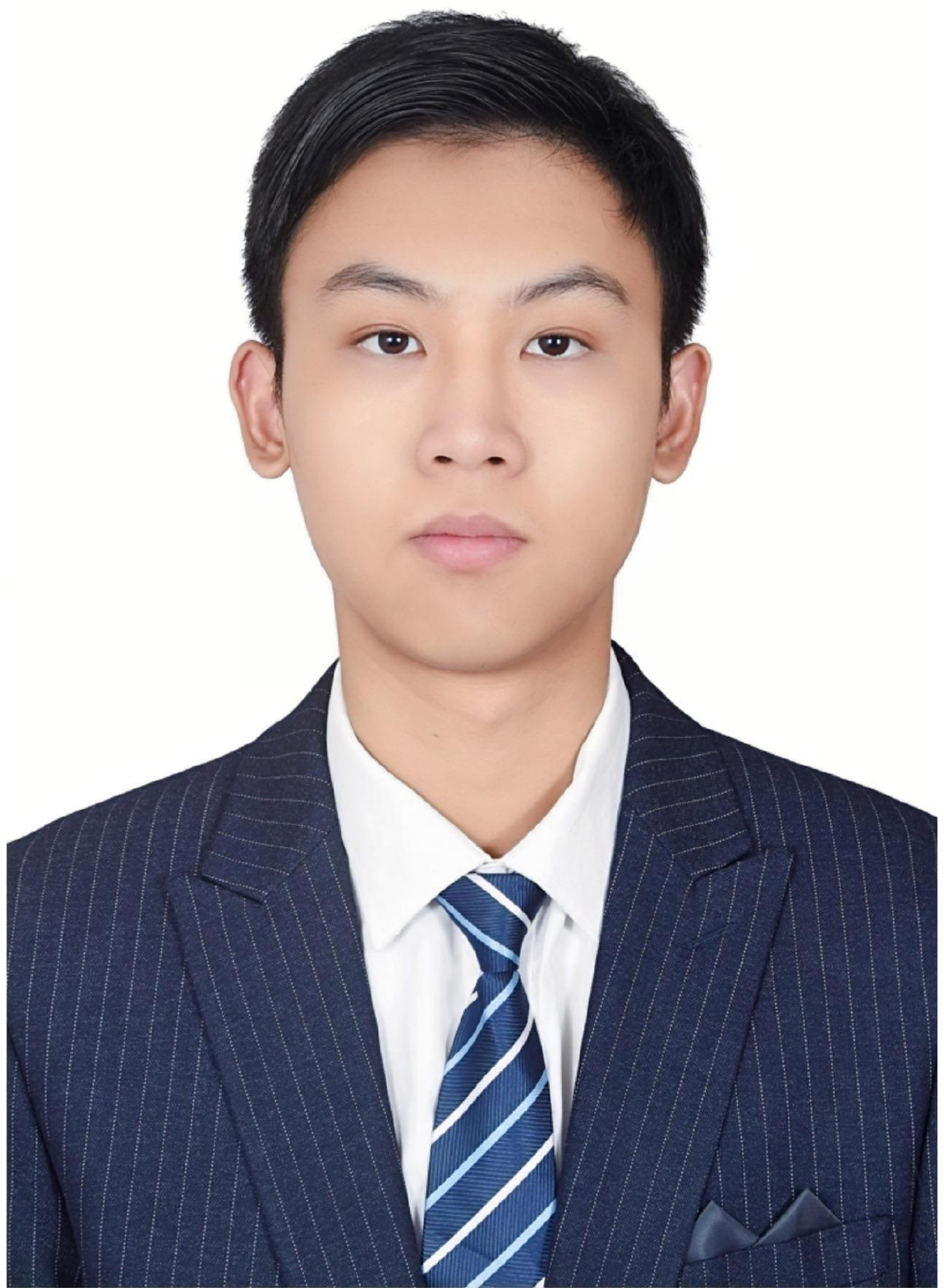}}] 
{Wentao Liang} received the B.S. degree from Guangdong University of Technology, Guangzhou, China, in 2025. He is currently pursuing an M.S. degree in the State Key Laboratory of Industrial Control Technology, College of Control Science and Engineering, Zhejiang University, Hangzhou, China. His research interests include multi-agent deep reinforcement learning, time series forecasting, and the computer vision.
\end{IEEEbiography}

\begin{IEEEbiography}
[{\includegraphics[width=1in,height=1.25in,clip,keepaspectratio]{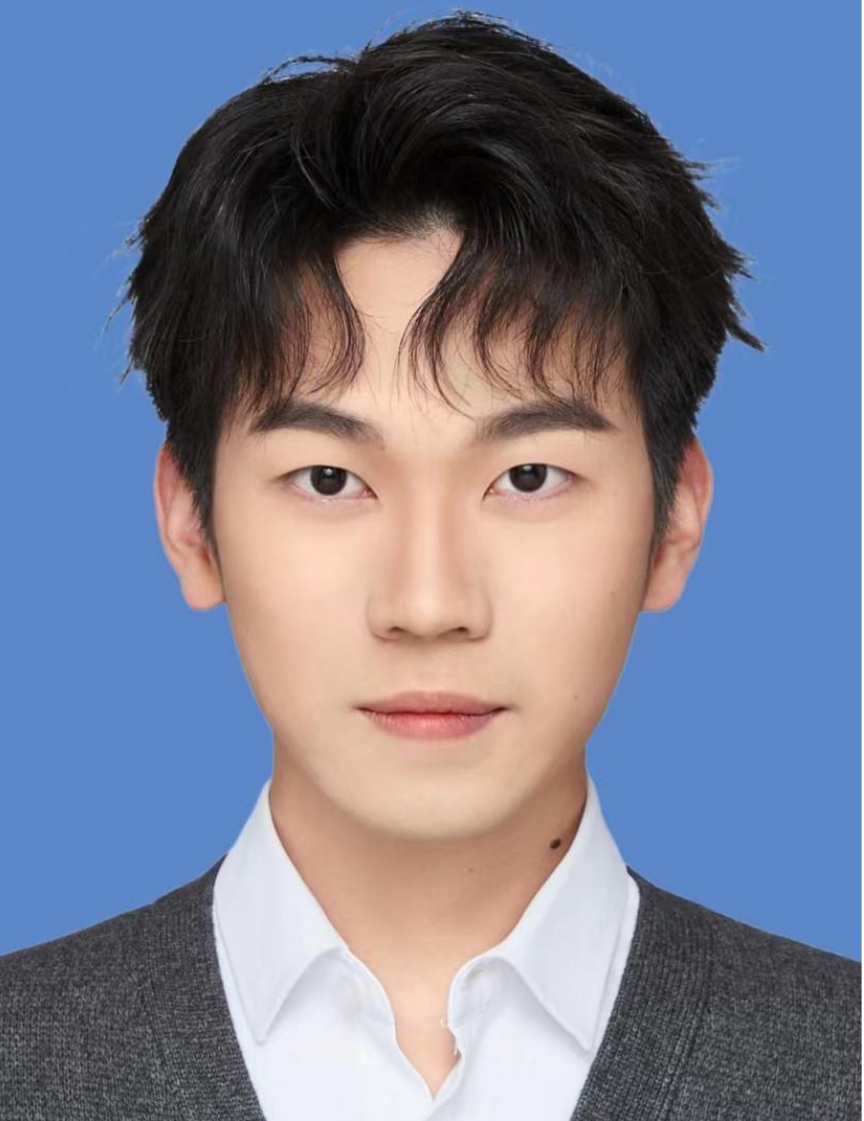}}]
{Jinbo Wen} received the B.Eng. degree from Guangdong University of Technology, China, in 2023. He is currently pursuing an M.S. degree with the College of Computer Science and Technology, Nanjing University of Aeronautics and Astronautics, China. His research interests include generative AI, blockchain, and the metaverse.
\end{IEEEbiography}

\begin{IEEEbiography}
[{\includegraphics[width=1in,height=1.25in,clip,keepaspectratio]{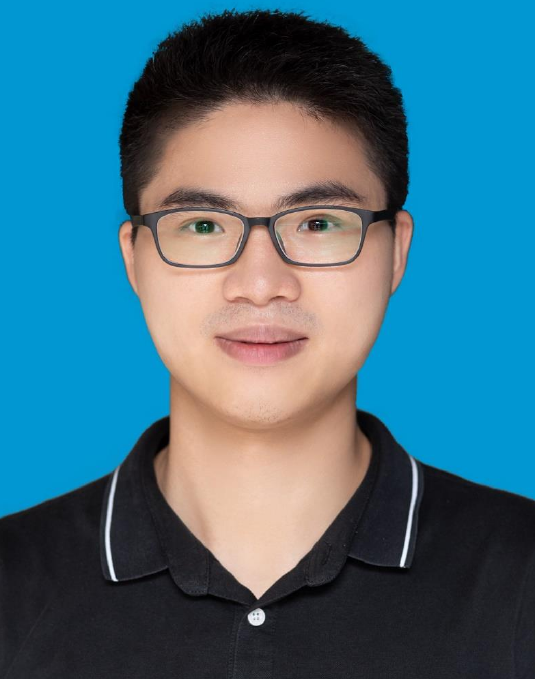}}]
{Jiawen Kang} (Senior Member, IEEE) received the Ph.D. degree from the Guangdong University of Technology, China, in 2018. He was a postdoc at Nanyang Technological University, Singapore, from 2018 to 2021. He is currently a professor at Guangdong University of Technology, China. His research interests mainly focus on blockchain, security, and privacy protection in wireless communications and networking.
\end{IEEEbiography}

\begin{IEEEbiography}
[{\includegraphics[width=1in,height=1.25in,clip,keepaspectratio]{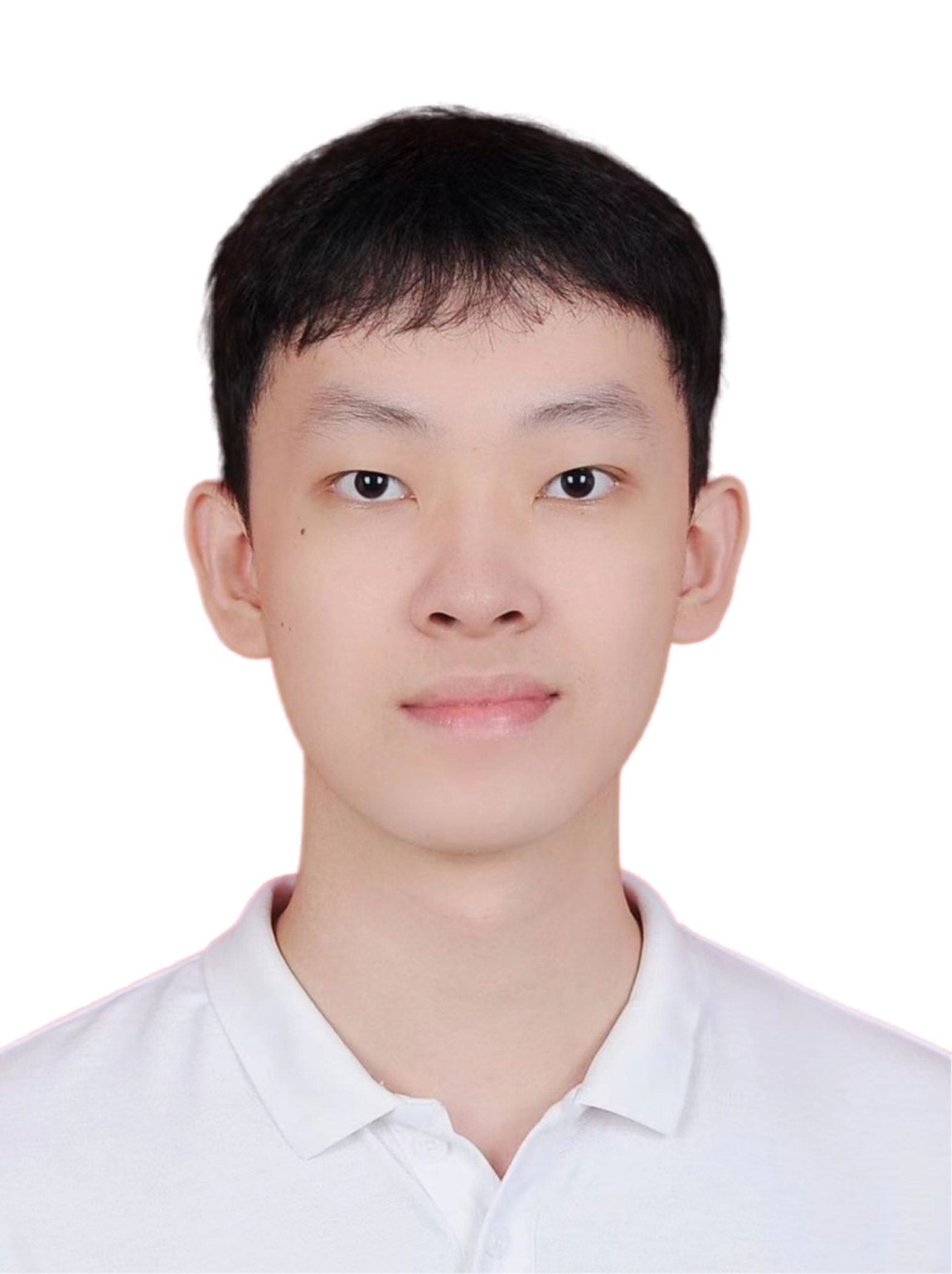}}]
{Junlong Chen} received the B.S. degree from Guangdong University of Technology, Guangzhou, China, in 2025. He is currently working toward the Ph.D. degree in Intelligent Transportation Thrust, The Hong Kong University of Science and Technology (Guangzhou), Guangzhou, China. His research interests mainly focus on Synesthesia of Machines, digital-twin, deep reinforcement learning, and AIGC in wireless communications and networking.
\end{IEEEbiography}

\begin{IEEEbiography}
[{\includegraphics[width=1in,height=1.25in,clip,keepaspectratio]{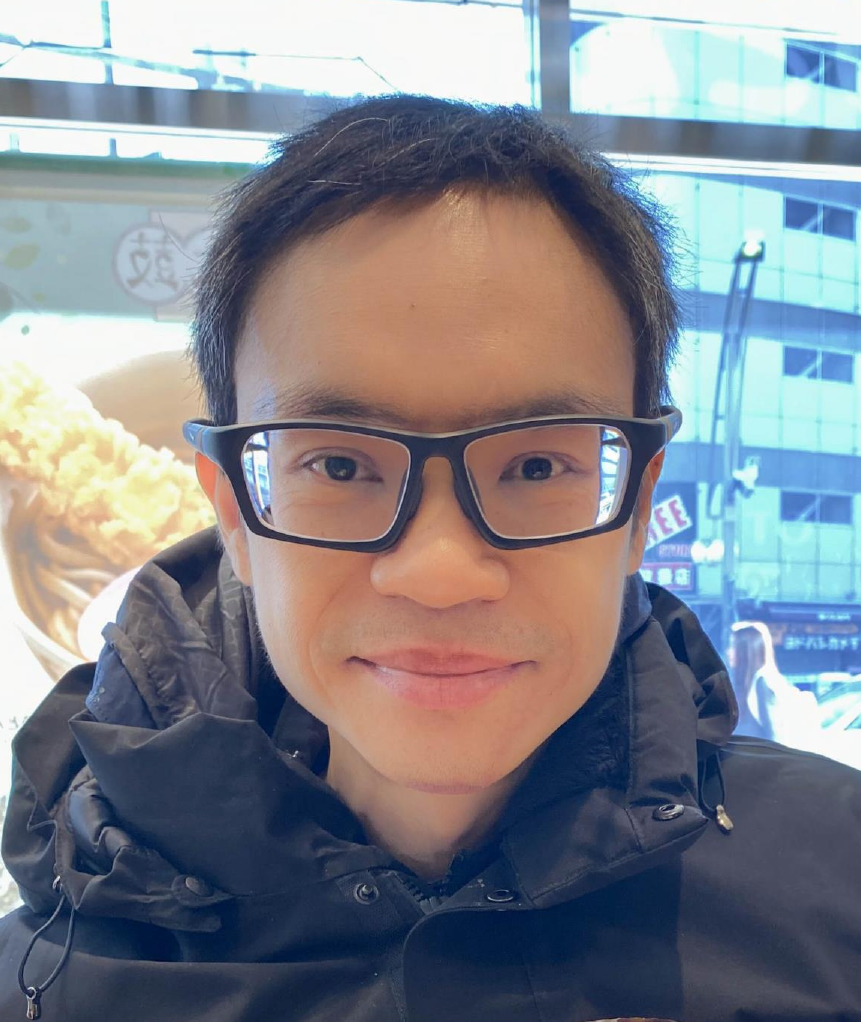}}] 
{Dusit Niyato} (Fellow, IEEE) is a professor in the School of Computer Science and Engineering at Nanyang Technological University, Singapore. He received a B.Eng. from King Mongkut's Institute of Technology Ladkrabang (KMITL), Thailand, in 1999 and a Ph.D. in Electrical and Computer Engineering from the University of Manitoba, Canada, in 2008. His research interests are in the areas of generative AI, edge intelligence, decentralized machine learning, and incentive mechanism design.
\end{IEEEbiography}

\end{document}